\documentclass[prd,%preprint,%twocolumn,
tightenlines,superscriptaddress,showpacs,
nofootinbib,eqsecnum,amsfonts,amsmath]{revtex4}
\usepackage{bm,graphics,graphicx,epsfig,color,ulem}
\allowdisplaybreaks
\def\be{\begin{equation}}
\def\ee{\end{equation}}
\def\ba{\begin{eqnarray}}
\def\ea{\end{eqnarray}}

\def\la{\langle}
\def\ra{\rangle}
\def\bse{\begin{subequations}}
\def\ese{\end{subequations}}
\def\rl{\right.\nonumber\\&\left.}
\def\rrll{\right.\right.\nonumber\\&\left.\left.}
\def\rrrlll{\right.\right.\right.\nonumber\\&\left.\left.\left.}
\def\rrrrllll{\right.\right.\right.\right.\nonumber\\&\left.\left.\left.\left.}

\begin{document}
%%%%%%%%%%%%%%%%%%%%%%%%%%%%%%%%%%%%%%%%%%%%%%TITLE%%%%%%%%%%%%%%%%%%%%%%%%%%%%%%%%%%%%%%%%%%%%%%%%
\title{Radial infall of two compact objects: 2.5PN linear momentum flux and associated recoil}
\author{Chandra Kant Mishra}\email{chandra@rri.res.in}
\affiliation{Raman Research Institute, Bangalore 560 080, India}
\affiliation{Indian Institute of Science, Bangalore 560 012, India}
%%%%%%%%%%%%%%%%%%%%%%%%%%%%%%%%%%%%%%%%%%%%%%ABSTRACT%%%%%%%%%%%%%%%%%%%%%%%%%%%%%%%%%%%%%%%%%%%%%
\begin{abstract}
The loss rate of linear momentum from a binary system composed of compact objects (radially falling 
towards each other under mutual gravitational influence) has been investigated using the multipolar 
post-Minkowskian approach. The 2.5PN accurate analytical formula for the linear momentum flux
is provided, in terms of the separation of the two objects, in harmonic coordinates, both for a 
finite and infinite initial separation. 
The 2.5PN formulas for the linear momentum flux are finally used to estimate the recoil velocity 
accumulated during a premerger phase of the binary evolution.
\end{abstract}
\date{\today}
\maketitle
%%%%%%%%%%%%%%%%%%%%%%%%%%%%%%%%%%%%%%%%%%%%%%INTRODUCTION%%%%%%%%%%%%%%%%%%%%%%%%%%%%%%%%%%%%%%%%%
\section{Introduction}
\label{sec:intro}
%++++++++++++++++++++++++++++++++++++++++++++++++++++++++++++++++++++++++++++++++++++++++++++++++++
%++++++++++++++++++++++++++++++++++++++++++++++++++++++++++++++++++++++++++++++++++++++++++++++++++
Gravitational waves from coalescing binary systems carry away energy 
and angular momentum of the source. For asymmetric binaries (composed of objects of unequal 
masses and/or with nonzero spins), there will also be a net loss of the linear momentum from the source. As 
a consequence, the center-of-mass of the source will receive a recoil in the opposite direction. 
This recoil accumulates until the two objects of the binary merge to form a single object and the 
source stops losing linear momentum. At this juncture, the remnant of the coalesced binary moves 
with a non zero kick speed along a straight line path in space. For a more detailed discussion on 
the phenomenon of gravitational wave recoil, see Ref.~\cite{2005gbha.conf..333H}.
The phenomenon of gravitational wave recoil is extremely important in various astrophysical contexts 
such as the formation 
and growth of super massive massive black holes at the centers of galaxies. If the recoil velocity 
of the remnant of the coalesced binary is more 
than its escape velocity from the host, then the host will not be able to retain the 
remnant and models that grow the super massive black holes via successive mergers from other 
black holes will not be favored \cite{Merritt:2004xa}. An accurate estimate for the recoil 
velocities associated with compact binary mergers can be used to address issues like observations 
of super massive black holes at the centers of most of the galaxies in the local universe 
\cite{Richstone98} or their apparent absence in globular clusters and dwarf galaxies or to predict 
the population of compact binary systems in globular clusters.

The importance of this phenomenon has been realized widely in astrophysics community and 
there have been numerous analytical or semi-analytical \cite{Fitchett83, Wi92, BQW05, K95, 
Racine:2008kj, Favata:2004wz, BuonD98,D01, Damour:2006tr, Sopuerta:2006et, Sopuerta:2006wj, 
LeTiec:2009yg, 2012PhRvD..85d4021M} and numerical studies \cite{Campanelli:2004zw, Baker:2006vn, 
Herrmann07, Gonzalez:2008bi, Herrmann:2007ac, Koppitz:2007ev, Campanelli:2007ew, Gonzalez:2007hi} 
to compute this effect. All these studies compute the recoil effects due to the loss of linear momentum 
from compact binary systems (which either have mass-asymmetry and/or have non zero spin) moving in 
quasi-Keplerian or in quasi-circular orbits. Numerical simulations for nonspinning black hole binaries 
moving in quasi-circular orbit \cite{Campanelli:2004zw, Baker:2006vn, Herrmann07, Gonzalez:2008bi} 	
have shown that the recoil velocity can be of the order of few hundred ${\rm km\,s}^{-1}$ while for 
spinning case \cite{Herrmann:2007ac, Koppitz:2007ev, Campanelli:2007ew, Gonzalez:2007hi} the recoil 
velocity estimates can reach up to few thousand ${\rm km\,s}^{-1}$. 

Although, head-on infall and the subsequent merger of two compact objects due to 
gravitational wave radiation reaction effects would be an insignificant astrophysical possibility, 
nevertheless it has been studied extensively using various analytical/numerical approaches. The motivation behind 
such a study is many-fold. To start with, due to the axial symmetry of the system, the two-dimensional 
problem of compact binary motion becomes one-dimensional and hence the treatment becomes simple. 
This also can act as a toy problem for comparing various analytical and numerical approaches in their 
most simplified versions. In addition to this, head-on collision can be considered as an approximation 
to the merger phase of the inspiralling compact binary evolution. Finally, as pointed out in 
\cite{Choi:2007eu}, head-on collision studies can be used to remove the uncertainties in the direction 
of the recoil of the remnant.
%++++++++++++++++++++++++++++++++++++++++++++++++++++++++++++++++++++++++++++++++++++++++++++++++++
%++++++++++++++++++++++++++++++++++++++++++++++++++++++++++++++++++++++++++++++++++++++++++++++++++

One of the earliest attempts to compute recoil effects due to the radial plunge of a test particle 
into a Schwarzschild black hole is due to Nakamura and Haugan \cite{Nakamura:1983hk} using the black 
hole perturbation theory. Using a close limit approximation method Andrade and Price
\cite{Andrade:1996pc} first computed the recoil effects due to head-on collision of two black holes.
On the numerical relativity front, Anninos and Brandt \cite{Anninos:1998wt} computed the recoil 
velocity due to head-on collision of two unequal mass black holes. Some other (relatively recent) 
analytical/numerical works \cite{Choi:2007eu, Nichols:2011ih, Lovelace:2009dg} compute the recoil 
effects taking in to account the asymmetry in mass and/or in the spin.
As far as PN calculations are concerned, although, the recoil effects in a head-on collision case
have not been investigated explicitly, one can use expressions for the linear momentum flux from 
nonspinning inspiralling compact binary systems moving in general orbits 
\cite{Fitchett83, Wi92, Racine:2008kj} to write equivalent expressions for the head-on case by 
using the following transformations 
\footnote{These transformations assume the motion in along the z-axis.} 
(as suggested in \cite{SPW95, Mishra:2010in}): 
\begin{eqnarray}
{\bf x}&=&z\,\hat{\bf n},\,\,\, {\bf v}=\dot{z}\,\hat{\bf n},\,\,\,r=z,\,\,\, v=\dot{r} = \dot{z}. 
\label{GOtoHOC}
\end{eqnarray}
Here, $z$ is the separation between the two objects (under radial infall) at a given instant and 
$\dot {z}$ is the first time derivative of $z$, giving the relative speed of objects at that 
instant. The most recent related PN work \cite{Racine:2008kj} gives 2PN accurate expressions 
for the instantaneous part of the linear momentum and hence one can use the above transformations to 
write the 2PN expression for the instantaneous part of the linear momentum flux in terms of $z$ and $\dot{z}$. 
In the present work, we not only calculate the instantaneous part of the flux explicitly for the head on 
case to a higher order (2.5PN as compared to previous 2PN calculations) but also compute additional 
terms contributing at the 1.5PN order and 2.5PN order (tail contribution) whose nature has been 
discussed in more detail in the next section.
 
%++++++++++++++++++++++++++++++++++++++++++++++++++++++++++++++++++++++++++++++++++++++++++++++++++
%++++++++++++++++++++++++++++++++++++++++++++++++++++++++++++++++++++++++++++++++++++++++++++++++++
In the present work, we compute the 2.5PN accurate analytical expressions for the linear momentum 
flux, in harmonic coordinates, emitted during the radial infall of two nonspinning compact objects 
under mutual gravitational influence. We study the problem for two different situations based on 
the initial separation between the two objects. In the first case we assume that initially the 
objects are separated by some finite distance ( we call it case (a)) and in the other case we 
assume that the initial separation between them is infinite (we call it case (b)). Linear momentum flux as 
a function of the separation between the two objects at any instant of time for the two situations, 
case (a) and case (b), are given by Eq.~\eqref{totallmf-FS} and \eqref{totallmf-IS}, respectively. 
We use these results to estimate the associated recoil velocity for the two situations. Since linear 
momentum flux expression (Eq.~\eqref{totallmf-FS}) involves some integrals (Eq.~\eqref{numints}) which 
can only be evaluated numerically, it is not possible to give analytical PN expressions for the 
accumulated recoil velocity for case (a) and thus has been computed numerically. However, for 
case (b), a 2.5PN accurate expression for the recoil velocity is given by Eq.~\eqref{recvel-IS}. 
A graphical representation of our results have been given in 
Figs.~\ref{fig:recvel-vsnu-vsgf}-\ref{fig:recvelvsnu-nPN}. We find that the recoil velocity is maximum 
for a binary with $\nu\sim 0.19$ and is of the order of $\sim$1.6${\rm km\,s}^{-1}$ if we terminate our 
calculations when the two objects are 5 $G\,m/c^2$ apart.

%++++++++++++++++++++++++++++++++++++++++++++++++++++++++++++++++++++++++++++++++++++++++++++++++++
This paper is organized in the following manner. In Sec.~\ref{sec:structure-lmf}, we first write the 
general formula for the linear momentum flux in terms of the radiative multipole moments 
of an isolated post-Newtonian source. Next, we use relations connecting the radiative multipole 
moments to the source multipole moments, to express the linear momentum flux in terms of the source 
multipole moments. Section~\ref{sec:inputs} lists all the inputs that will be required for 
computing the 2.5PN accurate analytical expression for the linear momentum flux. In Sec.~\ref{sec:lmfhoc}, 
we present the 2.5PN accurate analytical results for the linear momentum flux, in harmonic 
coordinates, for two situations (case (a) and case (b)). In Sec.~\ref{sec:recvel}, we show how 
the expressions for the linear momentum flux can be used to compute the associated recoil velocity 
accumulated till any epoch of the binary's evolution (within the validity of PN approximations). 
Finally, in Sec.~\ref{sec:results}, we summarize our findings and discuss the numerical estimates 
for the recoil velocity in the head-on case.
%%%%%%%%%%%%%%%%%%%%%%%%%%%%%%%%%%%%%%%%%%%%%%BasicFormulae%%%%%%%%%%%%%%%%%%%%%%%%%%%%%%%%%%%%%%%%
\section{The post-Newtonian Structure for the flux of linear momentum: Head-on Case}
\label{sec:structure-lmf}
%++++++++++++++++++++++++++++++++++++++++++++++++++++++++++++++++++++++++++++++++++++++++++++++++++
The general formula for linear momentum flux, in the far-zone of an isolated source, in terms 
of two sets of symmetric trace-free {\it radiative} multipole moments ($U_L, V_L$), is given in \cite{Th80} 
(see Eq.~(4.20\'\,) there). The {\it radiative} moments, $U_L(U)$ and $V_L(U)$, are referred as mass-type and 
current-type {\it radiative} multipole moments, respectively, and are functions of the retarded 
time $U$ in {\it radiative} coordinates. Here, $L=i_1i_2\cdots i_l$ represents a multi-index 
comprised of $l$ spatial indicies and $U$ is given by $U=T-R/c$, where $T$ and $R$ denote time of 
observation and the distance to the source in {\it radiative} coordinates, respectively. At 2.5PN 
order, the expression for linear momentum flux, in terms of radiative multipole moments 
($U_L$, $V_L$), reads 
\ba
\label{structure-lmf}
{\mathcal{F}_{P}^{i}}(U)&=& \frac{G}{c^7}\,\biggl\{\left[\frac{2}{63}\,
U^{(1)}_{ijk}\,U^{(1)}_{jk}+\frac{16}{45}\,\varepsilon_{ijk}
U^{(1)}_{ja}\,V^{(1)}_{ka}\right]\nonumber\\&&
+{1\over c^2}\left[\frac{1}{1134}\,U^{(1)}_{ijkl}\,U^{(1)}_{jkl}
+\frac{1}{126}\,\varepsilon_{ijk}U^{(1)}_{jab}\,V^{(1)}_{kab}
+\frac{4}{63}\,V^{(1)}_{ijk}\,V^{(1)}_{jk}\right]\nonumber\\&&
+{1\over c^4}\left[\frac{1}{59400}\,U^{(1)}_{ijklm}\,U^{(1)}_{jklm}
+\frac{2}{14175}\,\varepsilon_{ijk}U^{(1)}_{jabc}\,V^{(1)}_{kabc}
+\frac{2}{945}\,V^{(1)}_{ijkl}\,V^{(1)}_{jkl}\right]+{\cal O}\left({1\over c^6}\right) \biggr\}.
\ea
In the above, $\left\{U^{(1)}_L, V^{(1)}_L\right\}$, denote the $1^{st}$ time derivative of 
$\left\{U_L, V_{L}\right\}$, $\epsilon_{ijk}$ denotes the Levi-Civita tensor with $\epsilon_{123}=+1$ 
and ${\cal O}\left({1/c^6}\right)$ indicates that corrections of the order 3PN and above have been 
neglected in the present analysis. The expression for linear momentum flux, in terms of {\it radiative} 
multipole moments $(U_L, V_L)$, is not very useful unless we show how these moments are connected 
to the actual parameters of the source. Fortunately, the formalism for connecting {\it radiative} 
multipole moments to the source-rooted moments, with the PN accuracy desired in this work, has already 
been developed \cite{BFIS08} using the multipolar post-Minkowskian approach \cite{BFeom, Bliving, BIJ02, 
BI04mult, BDE04, BDEI05}. In the multipolar post-Minkowskian formalism, $U_L$ and $V_L$ are first written 
in terms of two sets of multipole moments, $M_L$ and $S_L$, referred as mass-type and current-type 
{\it canonical} multipole moments, respectively. Next, these {\it canonical} multipole moments, $M_L$ and $S_L$, 
are written in terms of six sets of multipole moments, $I_L, J_L, W_L, X_L, Y_L, Z_L$, referred as source 
multipole moments. The multipole moments, $I_L$ and $J_L$, thoroughly describe the source and are referred 
as mass-type and current type source multipole moments. The other four, $W_L$, $X_L$, $Y_L$ and $Z_L$ are
referred as gauge moments as they do not play any role in a linearized theory and only become important 
at nonlinear level. Reference \cite{BFIS08} explicitly lists all the relations connecting ($U_L$, $V_L$) 
to ($M_L$, $S_L$) (see Eqs.~(5.4)-(5.8) there) and those connecting ($M_L$, $S_L$) to ($I_L, \cdots , Z_L$) 
(see Eqs.~(5.9)-(5.11) there). Using these relations one can explicitly write expressions for {\it radiative} 
multipole moments ($U_L$, $V_L$) (and hence the linear momentum flux at 2.5PN order given by 
Eq.~\eqref{structure-lmf}) in terms of source multipole moments ($I_L\cdots Z_L$). Before we express radiative 
multipole moments in terms of source multipole moments, we would like to bring in to the notice the fact 
that, for head-on case current-type moments ($V_L$ or $S_L$ or $J_L$) would not contribute as they are 
proportional to the angular momentum, $\cal J$, which vanishes for the head-on case. This allows us to 
re-write Eq.~\eqref{structure-lmf}, in a form specific to a head-on case, and it reads
\ba
\label{structure-lmfhoc}
{\mathcal{F}_{P}^{i}}(U)&=& \frac{G}{c^7}\,\biggl\{
\frac{2}{63}\,U^{(1)}_{ijk}\,U^{(1)}_{jk}
+{1\over c^2}\left[\frac{1}{1134}\,U^{(1)}_{ijkl}\,U^{(1)}_{jkl}\right]
+{1\over c^4}\left[\frac{1}{59400}\,U^{(1)}_{ijklm}\,U^{(1)}_{jklm}\right]
+{\cal O}\left({1\over c^6}\right) \biggr\}.
\ea
It is evident from the above, that moments appearing at the lowest order in the PN series need to be 
known with the highest PN accuracy whereas those appearing at a higher PN order need to be known with 
smaller PN accuracy, {\it e.g.} in the present case we need $U_{ij}$ and $U_{ijk}$ to 2.5PN accuracy 
whereas $U_{ijkl}$ and $U_{ijklm}$ need to be known with 1.5PN and Newtonian accuracy, respectively. 
Now, making use of Eqs.~(5.4)-(5.7) and Eqs.~(5.9)-(5.11) of \cite{BFIS08} and keeping in mind that current 
type moments vanish for the head-on case, we write $U_L$ in terms of source multipole moments in  a 
form specific to the head-on case, which read
\bse\label{UL}
%++++++++++++++++++++++++++++++++++++++++++++++++++++++++++++++++++++++++++++++++++++++++++++++++++++
\begin{align}\label{U2}
U_{ij}(U) &= I^{(2)}_{ij}(U)+{2 G M\over c^3} \int_{0}^{\infty} d
\tau \left[\ln \left({c\tau\over 2 r_0}\right)+{11\over12} \right]
I^{(4)}_{ij} (U-\tau) \nonumber \\ &
+{G\over c^5}\left\{-{2\over7}\int_{0}^{\infty} d\tau I^{(3)}_{a\langle
i}(U-\tau)I^{(3)}_{j\rangle a}(U-\tau)+ {1\over7}I^{(5)}_{a\langle i}I_{j\rangle a} 
- {5 \over7} I^{(4)}_{a\langle i}I^{(1)}_{j\rangle a} 
-{2 \over7} I^{(3)}_{a\langle i}I^{(2)}_{j\rangle a}
\rl+4\left[W^{(2)}I_{ij}-W^{(1)}I_{ij}^{(1)}
\right]^{(2)}\right\}+\,\, \mathcal{O}\left(\frac{1}{c^6}\right),
\end{align}
%++++++++++++++++++++++++++++++++++++++++++++++++++++++++++++++++++++++++++++++++++++++++++++++++++++
\begin{align}\label{U3}
U_{ijk} (U) &= I^{(3)}_{ijk} (U) + {2G M\over c^3} \int_{0}^{\infty}
d\tau\left[ \ln \left({c \tau\over 2 r_0}\right)+{97\over60} \right]
I^{(5)}_{ijk} (U-\tau)\nonumber \\ & 
+{G\over c^5}\left\{-{1\over3}\int_{0}^{\infty}d\tau\,I^{(3)}_{a\langle i}(U-\tau)I^{(4)}_{jk\rangle 
a}(U-\tau)-{4\over3}I^{(3)}_{a\langle i}I^{(3)}_{jk\rangle a}
-{9\over4}I^{(4)}_{a\langle i}I^{(2)}_{jk\rangle a}
+{1\over4}I^{(2)}_{a\langle i}I^{(4)}_{jk\rangle a} \rl-
{3\over4}I^{(5)}_{a\langle i}I^{(1)}_{jk\rangle a} +{1\over4}I^{(1)}_{a\langle
i}I^{(5)}_{jk\rangle a}
+ {1\over12}I^{(6)}_{a\langle
i}I_{jk\rangle a} 
+{1\over4}I_{a\langle i}I^{(6)}_{jk\rangle a}
\rl+4\left[W^{(2)} I_{ijk}-W^{(1)} I_{ijk}^{(1)}
+3\, I_{\langle ij}Y_{k\rangle}^{(1)}\right]^{(3)}
\right\}+\mathcal{O}\left(\frac{1}{c^6}\right),
\end{align}
%++++++++++++++++++++++++++++++++++++++++++++++++++++++++++++++++++++++++++++++++++++++++++++++++++
\begin{align}\label{U4}
U_{ijkl}(U) &= I^{(4)}_{ijkl} (U) + {G\over c^3} \left\{ 2 M
\int_{0}^{\infty} d \tau \left[ \ln \left({c \tau\over
2 r_0}\right)+{59\over30} \right] I^{(6)}_{ijkl}(U-\tau) \right.\nonumber\\
&\quad\left. +{2\over5}\int_{0}^{\infty} d\tau I^{(3)}_{\langle
ij}(U-\tau)I^{(3)}_{kl\rangle }(U-\tau) -{21\over5}I^{(5)}_{\langle ij}I_{kl\rangle }
-{63\over5}I^{(4)}_{\langle ij}I^{(1)}_{kl\rangle }- {102\over5}I^{(3)}_{\langle
ij}I^{(2)}_{kl\rangle }\right\}
+\,\mathcal{O}\left(\frac{1}{c^5}\right),
\end{align}
%++++++++++++++++++++++++++++++++++++++++++++++++++++++++++++++++++++++++++++++++++++++++++++++++++
\begin{align}\label{U5}
U_{ijklm} (U) &= I^{(5)}_{ijklm}(U)+\mathcal{O}\left(\frac{1}{c^3}\right).
\end{align}
\ese
In the above, angular brackets ($\la\ra$) surrounding indices denote symmetric trace-free projections.
Here, $M$, is the total ADM mass of the source and $r_0$ is an arbitrary length scale and provides a 
scale for the logarithms in tail integrals. This length scale was first introduced in the multipolar 
post-Minkowskian formalism 
and enters the relation connecting the retarded time, $U$ in {\it radiative} coordinate to the retarded time, 
$u$=$t$-$r/c$ in {\it harmonic} coordinates, which reads
\be\label{U-u}
U=t-{r\over c}-{2 G M\over c^3}\ln\left({r\over r_0}\right)          
\ee
In addition, note the presence of two types of terms in above expressions: the first kind involves multipole moments
at any given retarded time $U$ and are referred as {\it instantaneous} terms and the other kind involves 
integrals over time, referred as {\it hereditary} terms that require the knowledge of multipole moments 
at any time $U'=U-\tau$ before $U$. Further, the {\it hereditary} terms can be split into two parts: 
terms with and without logarithmic factors inside the integrals. Integrals with logarithmic factor are called 
{\it tail} integrals and those without logarithmic factor are referred to as {\it memory} integrals.
%==================================================================================================

Since the linear momentum flux involves $1^{\rm st}$ time derivative of mass-type radiative multipole moments 
(Eq.~\eqref{structure-lmfhoc}), first we need to write $U_L^{(1)}$ in terms of source multipole moments.\footnote{
The {\it memory} integral is a time anti-derivative and thus becomes instantaneous 
when we take the time derivative of $U_L$.} In terms of source multipole moments, $U_L^{(1)}$ take the following form 
\bse
\label{UL1}
\begin{align}
U_{ij}^{(1)}(U)&=I_{ij}^{(3)}(U)
+{2 G M\over c^3} \int_{0}^{\infty} d
\tau \left[\ln \left({c\tau\over 2 r_0}\right)+{11\over12} \right]
I^{(5)}_{ij} (U-\tau)
+{G\over c^5}\left[{1\over7}\,I_{a\la i}^{(6)}I_{j\ra a}
-{4\over7}\,I_{a\la i}^{(5)}I_{j\ra a}^{(1)}
-I_{a\la i}^{(4)}I_{j\ra a}^{(2)}
-{4\over7}\,I_{a\la i}^{(3)}I_{j\ra a}^{(3)}
\rl+4\left[W^{(2)}I_{ij}-W^{(1)}I_{ij}^{(1)}\right]^{(3)}\right]
+{\cal O}\left({1\over c^6}\right) 
\label{U21},\\
U_{ijk}^{(1)}(U)&=I_{ijk}^{(4)}(U)
+ {2G M\over c^3} \int_{0}^{\infty}
d\tau\left[ \ln \left({c \tau\over 2 r_0}\right)+{97\over60} \right]
I^{(6)}_{ijk} (U-\tau)
+{G\over c^5}\left[-{43\over12}\,I_{a\la i}^{(4)}I_{jk\ra a}^{(3)}
-{17\over12}\,I_{a\la i}^{(3)}I_{jk\ra a}^{(4)}
-3\,I_{a\la i}^{(5)}I_{jk\ra a}^{(2)}
\rl+{1\over2}\,I_{a\la i}^{(2)}I_{jk\ra a}^{(5)}
-{2\over3}\,I_{a\la i}^{(6)}I_{jk\ra a}^{(1)}
+{1\over2}\,I_{a\la i}^{(1)}I_{jk\ra a}^{(6)}
+{1\over12}\,I_{a\la i}^{(7)}I_{jk\ra a}
+{1\over4}\,I_{a\la i}I_{jk\ra a}^{(7)}
+4\,\left[W^{(2)}I_{ijk}-W^{(1)}I_{ijk}^{(1)}+3\,I_{\la ij}Y_{k\ra}^{(1)}\right]^{(4)}\right]
\nonumber \\&+{\cal O}\left({1\over c^6}\right)\label{U31},\\
U_{ijkl}^{(1)}(U)&=I_{ijkl}^{(5)}(U)
+{G\over c^3}\left[2 M
\int_{0}^{\infty} d \tau \left[ \ln \left({c \tau\over
2 r_0}\right)+{59\over30} \right] I^{(7)}_{ijkl}(U-\tau)
-20\,I_{\la ij}^{(3)}I_{kl\ra}^{(3)}
-{84\over5}\,I_{\la ij}^{(5)}I_{kl\ra}^{(1)}
-33\,I_{\la ij}^{(4)}I_{kl\ra}^{(2)}
\rl-{21\over5}\,I_{\la ij}^{(6)}I_{kl\ra}
\right]
+{\cal O}\left({1\over c^5}\right) 
\label{U41},\\
U_{ijklm}^{(1)}(U)&=I_{ijklm}^{(6)}(U)+{\cal O}\left({1\over c^3}\right).
\label{U51}
\end{align}
\ese
It was argued and then shown in \cite{Mishra:2010in} (see Sec. II there for a detailed discussion) 
that the presence of $r_0$ in the {\it tail} integrals at 1.5PN order is due to our use of the 
{\it radiative} coordinates and will disappear if we insert $U$ (given by Eq.\eqref{U-u}) back 
in expressions for $U_L$ (same would be true for $U_L^{(1)}$). Upon doing so we can write expressions 
for $U_L^{(1)}$ in {\it harmonic} coordinates which now will be free from the arbitrary length scale, 
$r_0$, and read
\bse
\label{UL1har}
\begin{align}
U_{ij}^{(1)}(u)&=I_{ij}^{(3)}(u)
+{2 G M\over c^3} \int_{0}^{\infty} d
\tau \left[\ln \left({c\tau\over 2 r}\right)+{11\over12} \right]
I^{(5)}_{ij} (u-\tau)
+{G\over c^5}\left[{1\over7}\,I_{a\la i}^{(6)}I_{j\ra a}
-{4\over7}\,I_{a\la i}^{(5)}I_{j\ra a}^{(1)}
-I_{a\la i}^{(4)}I_{j\ra a}^{(2)}
-{4\over7}\,I_{a\la i}^{(3)}I_{j\ra a}^{(3)}
\rl+4\left[W^{(2)}I_{ij}-W^{(1)}I_{ij}^{(1)}\right]^{(3)}\right]
+{\cal O}\left({1\over c^6}\right) 
\label{U21har},\\
U_{ijk}^{(1)}(u)&=I_{ijk}^{(4)}(u)
+ {2G M\over c^3} \int_{0}^{\infty}
d\tau\left[ \ln \left({c \tau\over 2 r}\right)+{97\over60} \right]
I^{(6)}_{ijk} (u-\tau)
+{G\over c^5}\left[-{43\over12}\,I_{a\la i}^{(4)}I_{jk\ra a}^{(3)}
-{17\over12}\,I_{a\la i}^{(3)}I_{jk\ra a}^{(4)}
-3\,I_{a\la i}^{(5)}I_{jk\ra a}^{(2)}
\rl+{1\over2}\,I_{a\la i}^{(2)}I_{jk\ra a}^{(5)}
-{2\over3}\,I_{a\la i}^{(6)}I_{jk\ra a}^{(1)}
+{1\over2}\,I_{a\la i}^{(1)}I_{jk\ra a}^{(6)}
+{1\over12}\,I_{a\la i}^{(7)}I_{jk\ra a}
+{1\over4}\,I_{a\la i}I_{jk\ra a}^{(7)}
+4\,\left[W^{(2)}I_{ijk}-W^{(1)}I_{ijk}^{(1)}+3\,I_{\la ij}Y_{k\ra}^{(1)}\right]^{(4)}\right]
\nonumber \\&+{\cal O}\left({1\over c^6}\right)\label{U31har},\\
U_{ijkl}^{(1)}(u)&=I_{ijkl}^{(5)}(u)
+{G\over c^3}\left[2 M
\int_{0}^{\infty} d \tau \left[ \ln \left({c \tau\over
2 r}\right)+{59\over30} \right] I^{(7)}_{ijkl}(u-\tau)
-20\,I_{\la ij}^{(3)}I_{kl\ra}^{(3)}
-{84\over5}\,I_{\la ij}^{(5)}I_{kl\ra}^{(1)}
-33\,I_{\la ij}^{(4)}I_{kl\ra}^{(2)}
-{21\over5}\,I_{\la ij}^{(6)}I_{kl\ra}
\right]\nonumber \\
&+{\cal O}\left({1\over c^5}\right) 
\label{U41har},\\
U_{ijklm}^{(1)}(u)&=I_{ijklm}^{(6)}(u)+{\cal O}\left({1\over c^3}\right).
\label{U51har}
\end{align}
\ese
%==================================================================================================
Equation \eqref{UL1har} along with Eq.~\eqref{structure-lmfhoc} gives 2.5PN accurate expression 
for the linear momentum flux in terms of the source multipole moments in {\it harmonic} coordinates, 
in a form specific to the head-on case. Next, the resulting expression can be decomposed into two distinct 
pieces namely: the {\it instantaneous} contribution and the {\it hereditary} contribution whose nature 
has already been discussed above. The total linear momentum flux reads   
\begin{eqnarray}
{\mathcal{F}_{P}^{i}}&=&\left({\mathcal{F}_{P}^{i}}\right)_{\rm inst}
+\left({\mathcal{F}_{P}^{i}}\right)_{\rm hered}\,,
\label{decomposed-lmf}
\end{eqnarray}
where the instantaneous part is given by 
\begin{eqnarray}
\label{lmf-inst-IJ}
\left({\mathcal{F}_{P}^{i}}\right)_{\rm inst}&=&
\frac{G}{c^7}\,\left\{\frac{2}{63}\,I^{(4)}_{ijk}\,I^{(3)}_{jk}
+{1\over c^2}\left[\frac{1}{1134}\,I^{(5)}_{ijkl}\,I^{(4)}_{jkl}\right]
+{1\over c^4}\left[\frac{1}{59400}\,I^{(6)}_{ijklm}\,I^{(5)}_{jklm}\right]
\right.\nonumber\\&&\left.
+{G\over c^5}\left[{2\over63}\left(I_{ijk}^{(4)}\left[{1\over7}\,I_{a\la j}^{(6)}I_{k\ra a}
-{4\over7}\,I_{a\la j}^{(5)}I_{k\ra a}^{(1)}
-I_{a\la j}^{(4)}I_{k\ra a}^{(2)}
-{4\over7}\,I_{a\la j}^{(3)}I_{k\ra a}^{(3)}
+4\left[W^{(2)}I_{jk}-W^{(1)}I_{jk}^{(1)}\right]^{(3)}\right]
\right.\right.\right.\nonumber\\&&\left.\left.\left.
+I_{jk}^{(3)}\left[-{43\over12}\,I_{a\la i}^{(4)}I_{jk\ra a}^{(3)}
-{17\over12}\,I_{a\la i}^{(3)}I_{jk\ra a}^{(4)}
-3\,I_{a\la i}^{(5)}I_{jk\ra a}^{(2)}
+{1\over2}\,I_{a\la i}^{(2)}I_{jk\ra a}^{(5)}
-{2\over3}\,I_{a\la i}^{(6)}I_{jk\ra a}^{(1)}
+{1\over2}\,I_{a\la i}^{(1)}I_{jk\ra a}^{(6)}
\right.\right.\right.\right.\nonumber\\&&\left.\left.\left.\left.
+{1\over12}\,I_{a\la i}^{(7)}I_{jk\ra a}
+{1\over4}\,I_{a\la i}I_{jk\ra a}^{(7)}
+4\,\left[W^{(2)}I_{ijk}-W^{(1)}I_{ijk}^{(1)}
+3\,I_{\la ij}Y_{k\ra}^{(1)}\right]^{(4)}\right]\right)
+{1\over1134}\,I_{jkl}^{(4)}\left(-20\,I_{\la ij}^{(3)}I_{kl\ra}^{(3)}
\right.\right.\right.\nonumber\\&&\left.\left.\left.
-{84\over5}\,I_{\la ij}^{(5)}I_{kl\ra}^{(1)}
-33\,I_{\la ij}^{(4)}I_{kl\ra}^{(2)}
-{21\over5}\,I_{\la ij}^{(6)}I_{kl\ra}\right)\right]
+{\cal O}\left({1\over c^6}\right)\right\},\nonumber\\ 
\end{eqnarray}
%%%%%%%%%%%%%%%%%%%%%%%%%%%%%%%%%%%%%%%%%%%%%%%%%%%%%%%%%%%%%%%%%%%%%%%
where
\bse
\label{M2M3}
\begin{align}
\left[W^{(2)}I_{ij}-W^{(1)}I_{ij}^{(1)}\right]^{(3)} &=
\left[2\,W^{(4)}I_{ij}^{(1)}+W^{(5)}I_{ij}-W^{(1)}I_{ij}^{(4)}-2\,W^{(2)}I_{ij}^{(3)}\right]
\label{M2},\\
\left[W^{(2)} I_{ijk}-W^{(1)} I_{ijk}^{(1)}+3\, I_{\la ij}Y_{k\ra}^{(1)}\right]^{(4)} &=
\left[W^{(6)}I_{ijk}
+3\,W^{(5)}\,I_{ijk}^{(1)}
+2\,W^{(4)}\,I_{ijk}^{(2)}
-3\,W^{(2)}\,I_{ijk}^{(4)}
\right.\nonumber \\&\left.
-2\,W^{(3)}\,I_{ijk}^{(3)}
-W^{(1)}I_{ijk}^{(5)}
+3\,I_{\la ij}Y_{k \ra}^{(5)}
+12\,I_{\la ij}^{(1)}Y_{k \ra}^{(4)}
\right.\nonumber \\&\left.
+18\,I_{\la ij}^{(2)}Y_{k \ra}^{(3)}
+12\,I_{\la ij}^{(3)}Y_{k \ra}^{(2)}
+3\,I_{\la ij}^{(4)}Y_{k\ra}^{(1)}
\right]
\label{M3}.
\end{align}
\ese
%%%%%%%%%%%%%%%%%%%%%%%%%%%%%%%%%%%%%%%%%%%%%%%%%%%%%%%%%%%%%%%%%%%%%%%
and the hereditary contribution reads
\begin{eqnarray}
\label{lmf-hered-IJ}
{\left({\mathcal{F}_{P}^{i}}\right)_{\rm hered}}&=&
\frac{4\,G^2\,M}{63\,c^{10}}\,I_{ijk}^{(4)}(u)
\int_{0}^{\infty} d\tau \left[\ln \left({c \tau\over
2 r}\right)+{11\over12}\right]
I^{(5)}_{jk}(u-\tau)
\nonumber\\&&
+\frac{4\,G^2\,M}{63\,c^{10}}\,I_{jk}^{(3)}(u)\int_{0}^{\infty} d\tau \left[\ln
\left({c \tau\over2 r}\right)+{97\over60}\right]
I^{(6)}_{ijk}(u-\tau)
\nonumber\\&&
+{G^2\,M\over 567\,c^{12}}\,
I_{ijkl}^{(5)}(u)\int_{0}^{\infty} d\tau \left[\ln \left({c \tau\over
2 r}\right)+{97\over60} \right]
I^{(6)}_{jkl}(u-\tau)
\nonumber\\&&
+{G^2\,M\over 567\,c^{12}}\,I_{jkl}^{(4)}(u)\int_{0}^{\infty} d \tau
\left[\ln \left({c \tau\over 2 r}\right)+{59\over30} \right]
I^{(7)}_{ijkl}(u-\tau).
\end{eqnarray}
Now, if we know how the source multipole moments are related to the actual source parameters, 
with PN accuracy desired in the present work, and we have a suitable machinery to compute the 
time derivatives of the source multipole moments, we can express the linear momentum flux in 
terms of actual source parameters. With this motivation we move to our next section where 
we shall provide all necessary inputs that will be needed for computing the 2.5PN linear momentum 
flux in terms of the source parameters.    
%%%%%%%%%%%%%%%%%%%%%%%%%%%%%%%%%%%%%%%%%%%%%%%%%%%%%%%%%%%%%%%%%%%%%%%%%%%%%%%%%%%%%%%%%%%%%%%%%%%
\section{Inputs for computing the linear momentum flux: radial infall of two compact objects}
\label{sec:inputs} 
As discussed in Sec.~\ref{sec:intro}, in this paper we aim to study the loss rate of linear momentum 
(through outgoing gravitational waves) during the radial infall of two compact objects under mutual 
gravitational influence. Unlike the case of inspiralling compact binaries in eccentric or 
circular orbits (where the motion takes place in a plane), for the head-on case, the problem becomes 
one dimensional and thus the treatment becomes relatively simpler. For such sources, expressions 
connecting source multipole moments to the source parameters, with the PN accuracy desired in the 
present work, have been given in Ref.~\cite{Mishra:2010in}.\footnote{Reference \cite{Mishra:2010in} provides 
a 2PN expression for the mass octupole moment ($I_{ijk}$) however for the present purpose we need it 
with 2.5PN accuracy and this additional 2.5PN correction is new to this paper (see Eq.~\eqref{I3hoc}). In 
addition, the moment, $Y_i$, was not needed for the energy flux calculations at 3PN order but is needed here 
with Newtonian accuracy and is also new to this work (see Eq.~\eqref{dipoleY}).} Below we list all 
source multipole moments (in {\it harmonic} coordinates) needed for computing 2.5PN linear momentum 
flux in terms of the separation between the two objects at a given instant ($z$) and the first time 
derivative of $z$($\dot{z}$), giving the relative speed of objects at that instant (assuming the 
motion takes place along the z-axis).\footnote{Unlike Ref. \cite{Mishra:2010in}, where expressions 
for energy flux are given in standard harmonic (SH), 
modified harmonic (MH) and Arnowitt, Daser, and Misner (ADM) coordinates, here we only make use of 
{\it harmonic} coordinates for all relevant formulas. However, in the appendix we show how one can 
obtain equivalent analytical expressions for the linear momentum flux and recoil velocity in ADM 
coordinates.} The mass-type source multipole moments read
\bse\label{ILhoc}\begin{align}
\label{I2hoc}
I_{ij}&=\nu\,m\,z^2\left[1+\gamma\left(-\frac{5}{7}+\frac{8}{7}\nu\right)
+\gamma ^2 \left(-\frac{355}{252}-\frac{953}{126}\nu+\frac{337}{252}\nu^2\right)
+{\dot{z}^2\over c^2} \left(\frac{9}{14}-\frac{27}{14}\nu+\gamma\left(\frac{32}{9}
+\frac{289}{126}\nu-\frac{1195}{126}\nu^2\right)\right)
\rl+{\dot{z}^4\over c^4} \left(\frac{83}{168}-\frac{589}{168}\nu+\frac{1111}{168}\nu^2\right)
+\frac{24}{7} {\dot{z}\over c}\gamma^2\nu
\right]n_{\la ij\ra}+{\cal O}\left({1\over c^6}\right),\\
\label{I3hoc}
I_{ijk}&=-\nu\,m\,z^3\sqrt{1-4\,\nu}\left[1+\gamma  \left(-\frac{5}{6}+\frac{13}{6}\nu\right)
+\gamma ^2 \left(-\frac{47}{33}-\frac{1591}{132}\nu+\frac{235}{66}\nu^2\right)
+{\dot{z}^2\over c^2} \left(\frac{5}{6}-\frac{19}{6}\nu+\gamma  \left(\frac{54}{11}
+\frac{521}{132}\nu \rrrlll-\frac{2467}{132}\nu^2\right)\right)
+{\dot{z}^4\over c^4} \left(\frac{61}{88}-\frac{1579}{264}\nu+\frac{1129}{88}\nu^2\right)
+\frac{416}{45} {\dot{z}\over c} \gamma ^2 \nu
-\frac{12}{5} {\dot{z}^3\over c^3} \gamma  \nu
\right]n_{\la ijk\ra}+{\cal O}\left({1\over c^6}\right),\\
\label{I4hoc}
I_{ijkl}&=\nu\,m\,z^4\left[1-3 \nu +\gamma  \left(-\frac{10}{11}+\frac{61}{11}\nu
-\frac{105}{11}\nu^2\right)+{\dot{z}^2\over c^2} \left(\frac{23}{22}-\frac{159}{22}\nu
+\frac{291}{22}\nu^2\right)
\right]n_{\la ijkl\ra}+{\cal O}\left({1\over c^4}\right),\\
\label{I5hoc}
I_{ijklm}&=-\nu\,m\,z^5\sqrt{1-4\,\nu}\left(1-2\,\nu\right)n_{\la ijklm\ra}+{\cal O}\left({1\over c^2}\right).
\end{align}\ese 
Here, $n_i$ is the component of the unit vector, $\hat{n}$, along the direction of motion 
and $\gamma$ is our PN parameter and is related to the separation ($z$), between the objects 
at any instant of time, by $\gamma=(G m/c^2 z)$. In addition to this, one would also need 
1PN accurate expression for mass monopole (while computing hereditary terms), which can 
be identified with the ADM mass ($M$) of the system and Newtonian order expressions for 
gauge moments such as the one related to monopolar moment $W$ and dipolar moment $Y_i$ and 
are given as
\bse\label{M-W-Y}
\begin{align}\label{ADMmass}
M&=m\left(1-{\nu\over2}\gamma\right)+{\cal O}\left({1\over c^4}\right),\\
\label{monopoleW}
W&={1\over3}\nu m z\dot{z}+{\cal O}\left({1\over c^2}\right),\\
\label{dipoleY}
Y_i&={1\over5}\nu m z\sqrt{1-4\nu}\left({1\over 2}\frac{G m}{z}-\dot{z}^2\right)n_i
+{\cal O}\left({1\over c^2}\right).
\end{align}	
\ese
Having expressed, the source multipole moments in terms of the parameters of the source, 
now we need to compute relevant time derivatives of the source multipole moments. With 
mass-type source multipole moments and other required moments given in terms of $z$ 
and $\dot{z}$, whenever a time-derivative is taken, terms involving $\ddot{z}$ appear 
and thus one would need an expression for $\ddot{z}$ in terms of $z$ and $\dot{z}$ in 
order to write the linear momentum flux in terms of just $z$ and $\dot{z}$. 
Reference \cite{Mishra:2010in} lists somewhat general 3PN expression for $\ddot{z}$ (in terms 
of $z$ and $\dot{z}$) which can be used to write related expressions in SH, MH and ADM 
coordinates by choosing appropriate values for the parameters, $\alpha$ and $\beta$ 
(see Sec. IIIA of \cite{Mishra:2010in} for details). However, for our present purpose 
we just need 2.5PN accurate expressions for the $\ddot{z}$ in {\it harmonic coordinates} 
which can be obtained using $\alpha=-1$ and $\beta=0$ in Eq.~(3.5) of \cite{Mishra:2010in} 
and it reads\footnote{Note that at 2.5 PN order SH coordinates and MH coordinates are equivalent.}
\begin{align}\label{acc}
\ddot{z}&=-{G\,m\over z^2}\left[
1+\gamma  (-4-2 \nu )+\gamma ^2 \left(9+\frac{87}{4}\nu\right)
+{\dot{z}^2\over c^2} \left(-3+\frac{7}{2}\nu+\gamma  \left(-11 \nu +4 \nu ^2\right)\right)
\rl+{\dot{z}^4\over c^4} \left(-\frac{21}{8}\nu-\frac{21}{8}\nu^2\right)
-\frac{64}{15} {\dot{z}\over c} \gamma ^2 \nu
-\frac{16}{5} {\dot{z}^3\over c^3} \gamma  \nu
\right]+{\cal O}\left({1\over c^6}\right).
\end{align}
With, source multipole moments and $\ddot{z}$ expressed in terms of $z$ and $\dot{z}$, we 
can compute all relevant time-derivatives of source multipole moments appearing in flux 
formula (Eq.~\eqref{lmf-inst-IJ}-\eqref{lmf-hered-IJ}) and then can use them to write the 
linear momentum flux (at least instantaneous part of flux since hereditary contribution 
shall involve computing the integrals) in terms of $z$ and $\dot{z}$. However, following 
\cite{SPW95, Mishra:2010in}, we would like to write the expression for the linear 
momentum flux as a function of the separation of the two objects, 
alone. Also, we would like to compute the flux of linear momentum for two different 
situations: case (a) the two objects in the problem, initially separated by some finite distance, 
start falling radially from the rest, under mutual gravitational attraction, and case (b) a 
similar situation of radial infall but assumes infall from infinity. In order 
to write the linear momentum flux as a function of the separation of the two objects, 
we need an expression for $\dot{z}$ in terms of $z$, with a certain 
PN accuracy (here it should be 2.5PN accurate). In addition to this, $\dot{z}(z)$ 
is also sensitive to the initial conditions (case (a) and case (b)). At 3PN order, 
$\dot{z}(z)$ has been computed in \cite{Mishra:2010in} for the two different situations 
we want to explore in the present work and will not be reproduced here. We directly quote the result. 
In harmonic coordinates, 2.5PN expression for $\dot{z}$, in case of infall from a finite 
initial separation ($z_i$) is given as
\begin{align}\label{zdot-FS}
\dot{z}&=-\sqrt{2} c \sqrt{1-s} \sqrt{\gamma }\left[
1+\gamma  \left(-\frac{5}{2}+\frac{5}{4}\nu+s \left(\frac{1}{2}-\frac{9}{4}\nu\right)\right)
+\gamma ^2 \left(\frac{27}{8}-7 \nu +\frac{55}{32}\nu^2+s \left(-\frac{5}{4}+\frac{123}{8}\nu
-\frac{59}{16}\nu^2\right)\rrll+s^2 \left(\frac{3}{8}+\frac{\nu}{2}+\frac{47}{32}\nu^2\right)
\right)+\frac{8}{15} \sqrt{2} \sqrt{1-s} \gamma ^{5/2} \nu
+{\cal O}\left({1\over c^6}\right)
\right],
\end{align}
where, $s=z/z_i<1$.\footnote{Note that, the 2.5PN expression for $\dot{z}$ has been obtained by adding Eq.~(3.8)
and Eq.~(5.3) of \cite{Mishra:2010in} (as was suggested there) and then truncating resulting expression at the 2.5PN order.}
Related expression for the case of infall from infinity can be obtained by setting
$s=z/z_i$ in the above and then taking the limit as $z_i\rightarrow\infty$, and it reads
\begin{align}\label{zdot-IS}
\dot{z}&=-\sqrt{2} c \sqrt{\gamma }\left[
1+\gamma  \left(-\frac{5}{2}
+\frac{5}{4}\nu\right)
+\gamma ^2 \left(\frac{27}{8}-7 \nu +\frac{55}{32}\nu^2
\right)
+\frac{8}{15} \sqrt{2}\gamma ^{5/2} \nu
+{\cal O}\left({1\over c^6}\right)
\right].
\end{align} 
With these inputs we now are in a position to write the instantaneous part of the linear momentum 
flux in terms of the separation between the two objects under radial infall. However, the computation 
of hereditary contribution shall require 1PN expression for the trajectory of the 
problem.\footnote{Note that the leading order hereditary contribution occurs at 1.5PN order and thus  
computation of hereditary contribution at 2.5PN order shall only require 1PN inputs.} The 1PN 
trajectory for the two situations (case (a) and case (b)) have been given in \cite{Mishra:2010in}
(see Eq.~(3.23)-(3.24) and Eq.(3.26) there) and we simply recall it here (with slight change in 
presentation). For case (a), 
\be\label{1PNtraj-FS}
u={z_i^{3/2}\over \sqrt{2}\sqrt{G}\sqrt{m}}\left[g(s)-{1\over2}{G m\over c^2 z_i}
\left(h_0(s)-{\nu\over2} h_1(s) \right)\right]\ee
where $g(s)=f_1(s)-f_2(s)$, $h_0(s)=f_1(s)+9 f_2(s)$ and 
$h_1(s)=9 f_1(s)+f_2(s)$ with $f_1(s)=\sqrt{s}\sqrt{1-s}$ and $f_2(s)=\arcsin\sqrt{s}$.
For case (b), the above expression reduces to
\begin{equation}
\label{1PNtraj-IS}
u = -\frac{\sqrt{2}\, z^{3/2}}{3\, \sqrt{G}\, \sqrt{m}}\left[1+{15 \over 2}\frac{G\, m}{c^2\, z}
\left(1-\frac{\nu}{2}\right)\right]\,.
\end{equation}
We now have all the inputs to compute both the instantaneous and the hereditary 
contributions to the linear momentum flux, given by 
Eq.~\eqref{lmf-inst-IJ}-\eqref{lmf-hered-IJ}, and have been computed in the following 
section. 
%%%%%%%%%%%%%%%%%%%%%%%%%%%%%%%%%%%%%%%%%%%%%%Analytical Estimates%%%%%%%%%%%%%%%%%%%%%%%%%%%%%%%%%
\section{The 2.5PN linear momentum flux}
\label{sec:lmfhoc}
\subsection{The Instantaneous Contribution}
\label{subsec:lmfhoc-inst}
Instantaneous part of the linear momentum flux, in terms of the source multipole moments and their 
time derivatives, is given by Eq.~\eqref{lmf-inst-IJ}-\eqref{M2M3}. Expressions for the source
multipole moments (Eq.~\eqref{ILhoc}-\eqref{M-W-Y}) and the one for $\ddot{z}$ (Eq.~\eqref{acc}), 
in terms of $z$ and $\dot{z}$, can be used to compute the relevant time-derivatives of source 
multipole moments algebraically as functions of $z$ and $\dot{z}$. Next, in order to express the 
source multipole moments and their relevant time-derivatives, solely as functions of $z$, we need 
to make use of expression for $\dot{z}$ given in Eq.~\eqref{zdot-FS}-\eqref{zdot-IS}, depending upon 
the case we want to explore (case (a) or case (b)). Using, source multipole moments and their relevant 
time derivatives, solely expressed as functions of $z$, in Eq.~\eqref{lmf-inst-IJ}-\eqref{M2M3}, 
performing contraction of indices and truncating the resulting expression at 2.5PN order, we can write 
2.5PN accurate expression for the linear momentum flux as a function of separation of the two objects 
($z$).        
\subsubsection{Case (a): Infall from a finite distance }
\label{subsubsec:lmf-inst-FS}
The 2.5PN accurate expression for the linear momentum flux, for the situation which assumes the radial 
infall of two compact objects (initially separated by some finite distance $z_i$), in terms of our 
post-Newtonian parameter $\gamma$, reads  
\begin{align}\label{instlmf-FS}
\left({\mathcal{F}_{P}^{i}}\right)_{\rm inst}&=-{32\,\sqrt{2}\over105}{c^4\over G}\sqrt{1-s}\gamma^{11/2}\sqrt{1-4\nu}\nu^2
\left[s+\gamma\left(-\frac{425}{36}+\frac{25}{9}\nu
+s \left(-\frac{71}{18}+\frac{277}{36}\nu\right)
+s^2 \left(\frac{61}{6}-\frac{113}{12}\nu\right)\right)
\rl+\gamma ^2 \left(\frac{363379}{2376}-\frac{315163}{1584}\nu+\frac{14635}{396}\nu^2
+s \left(-\frac{99647}{594}+\frac{278611}{1584}\nu+\frac{12965}{3168}\nu^2\right)
+s^2 \left(-\frac{4801}{132}+\frac{125819}{792}\nu \rrrlll
-\frac{129959}{1584}\nu^2\right)+s^3 \left(\frac{7399}{264}-\frac{12527}{132}\nu
+\frac{13873}{352}\nu^2\right)\right)
+\frac{\gamma ^{5/2}\nu}{\sqrt{2} \sqrt{1-s}} \left(\frac{844}{45}-\frac{536}{15} s+\frac{1252}{45} s^2-\frac{464}{45} s^3\right)
\rl+{\cal O}\left({1\over c^6}\right)
\right]{n_i},
\end{align}
where $\gamma=(G m/c^2 z)$ and $s=z/z_i<1$. In the above, note that the leading order 
contribution to the linear momentum flux is proportional to the parameter $s$ and hence 
will vanishes for the case where initial separation is assumed to be infinite 
($z_i\rightarrow \infty$ {\it i.e.} $s\rightarrow0$). This is expected since the Newtonian 
order linear momentum flux is proportional to the $4^{th}$ time-derivative of the octupole 
moment ($I_{ijk}$), which vanishes for the case of infall from infinity.\footnote{This was 
first noted and discussed in \cite{Nakamura:1983hk} and can be verified easily.} However, 
for the case of infall from some finite separation the $I_{ijk}^{(4)}$ survives 
\cite{Mishra:2010in}, and hence we see a finite Newtonian order contribution to the linear 
momentum flux. 
\subsubsection{Case (b): Infall from infinity}
\label{subsubsec:lmf-inst-IS}
For the case of infall from infinity the related expression can be obtained by setting $s=z/z_i$
and then taking the limit as $z_i\rightarrow\infty$ we obtain
\begin{align}\label{instlmf-IS}
\left({\mathcal{F}_{P}^{i}}\right)_{\rm inst}&=-{32\,\sqrt{2}\over105}{c^4\over G}\gamma^{11/2}
\sqrt{1-4\nu}\nu^2 \left[\gamma\left(-\frac{425}{36}+\frac{25}{9}\nu\right)
+\gamma ^2 \left(\frac{363379}{2376}-\frac{315163}{1584}\nu+\frac{14635}{396}\nu^2\right)
+{422\sqrt{2}\over45}\gamma ^{5/2}\nu
\rl+{\cal O}\left({1\over c^6}\right)
\right]n_{i}.
\end{align}
\subsection{The Hereditary Contribution}
\label{subsec:lmf-hered}
The hereditary contribution to the linear momentum flux, in terms of time-derivatives of the source 
multipole moments, is given by Eq.~\eqref{lmf-hered-IJ}. Computing hereditary terms is relatively 
less easy as compared to computing instantaneous terms since it requires one to compute integrals 
over retarded time spanning over the entire dynamical history of the source. Now, since the leading 
order contribution to the linear momentum flux occurs at relative 1.5PN order we need to compute 
the hereditary effects only with relative 1PN accuracy in order to achieve relative 2.5PN accuracy 
for the present purpose. Moreover, only first two terms of Eq.~\eqref{lmf-hered-IJ} need to be 1PN 
accurate as the last two already contribute at 2.5PN order. In addition to this, in order to compute 
hereditary terms with accuracy desired in the present work, essentially we need to evaluate only 
three integrals, since integrals appearing in $2^{nd}$ and $3^{rd}$ term of Eq.~\eqref{lmf-hered-IJ} 
are essentially the same. Below,
we list the three integrals we need to evaluate (note $r\rightarrow z$)
\bse\label{heredint}
\begin{align}
\label{heredint1}
I_{1}&=\int_{0}^{\infty} d\tau \left[\ln \left({c\tau\over 2 z}\right)+{11\over12} \right]
I^{(5)}_{ij} (u-\tau),\\
\label{heredint2}
I_{2}&=\int_{0}^{\infty} d\tau \left[\ln \left({c\tau\over 2 z}\right)+{97\over60} \right]
I^{(6)}_{ijk} (u-\tau),\\
\label{heredint3}
I_{3}&=\int_{0}^{\infty} d\tau \left[\ln \left({c\tau\over 2 z}\right)+{59\over30} \right]
I^{(7)}_{ijkl} (u-\tau).
\end{align}
\ese
As discussed above, $I_1$ and $I_2$ need to be 1PN accurate whereas we need $I_3$ to be only 
Newtonian accurate.
\subsubsection{Case (a): Infall from a finite distance}
\label{subsubsec:lmf-hered-FS}
In this case, the integrals listed above can take the following form \cite{Mishra:2010in}
\bse\label{heredint-FS}
\begin{align}
\label{heredint1-FS}
I_{1}&=\int_{u(z_i)}^{u} d\tau \left[\ln \left({c\over 2 z}(u-\tau)\right)+{11\over12} \right]
I^{(5)}_{ij} (\tau),\\
\label{heredint2-FS}
I_{2}&=\int_{u(z_i)}^{u} d\tau \left[\ln \left({c\over 2 z}(u-\tau)\right)+{97\over60} \right]
I^{(6)}_{ijk} (\tau),\\
\label{heredint3-FS}
I_{3}&=\int_{u(z_i)}^{u} d\tau \left[\ln \left({c\over 2 z}(u-\tau)\right)+{59\over30} \right]
I^{(7)}_{ijkl} (\tau).
\end{align}
\ese
Note that for the infall from infinity case, when $z_i\rightarrow \infty$, $u(z_i)=u(\infty)=-\infty$.
With, required derivatives of the source multipole moments, expressed in terms of $z$, and the 
1PN trajectory (given by Eq. \eqref{1PNtraj-FS}) we can evaluate these integrals and they read
\bse\label{heredinteva}
\begin{align}
I_{1}&=
\frac{G^2 m^3  \nu}{z^4}\left\{
\frac{55}{6}-5 \ln(8 \gamma)
+s\left(-\frac{22}{3}+4 \ln(8 \gamma)\right)
+s^4 \left(-\frac{11}{6}+2 \,{\rm Int1}(s)+\ln(8 \gamma)\right)
\rl
+\gamma  \left[
-\frac{187}{3}(1-\nu)+34(1- \nu ) \ln(8 \gamma)
+s \left(\frac{209}{2}-\frac{737}{6}\nu+(-57+67 \nu ) \ln(8 \gamma)\right)
+s^2\left(-\frac{880}{21}+\frac{1177}{21}\nu
\rrrlll+\left(\frac{160}{7}-\frac{214}{7}\nu\right) \ln(8 \gamma)\right)
+s^5 \left(-\frac{11}{42}+\frac{187}{42}\nu-2 \,{\rm Int20}(s)+2 \nu 
\,{\rm Int21}(s)+\,{\rm Int30}(s)-\nu  \,{\rm Int31}(s)-\,{\rm Int4}(s)
\rrrlll+\frac{1}{2} \nu  \,{\rm Int5}(s)+\left(\frac{1}{7}-\frac{17\nu }{7}\right) \ln(8 \gamma)\right)
\right]\right\}n_{\la i}n_{j \ra},\\
%%%%%%%%%%%%%
I_2&=\frac{G^{5/2} m^{7/2} \nu }{z^{9/2}} \sqrt{1-4\nu} \left\{
\sqrt{2} \sqrt{1-s}\,s \left(-\frac{194}{5}+12 \ln(8 \gamma)\right)
-12 \sqrt{2} s^{9/2} \,{\rm Int6}(s)
\rl+\gamma  \left[
\sqrt{2} \sqrt{1-s} \left(\frac{8245}{9}-\frac{1940}{9}\nu+\left(-\frac{850}{3}
+\frac{200}{3}\nu\right) \ln(8 \gamma)
+s \left(\frac{388}{3}-\frac{1261}{6}\nu
+(-40+65 \nu ) \ln(8 \gamma)\right)
\rrrlll+s^2 \left(-\frac{679}{3}+\frac{8633}{30}\nu
+(70-89 \nu ) \ln(8 \gamma)\right)\right)
+s^{11/2} \left(-3 \sqrt{2} \nu  \,{\rm Int10}(s)
+12 \sqrt{2} \,{\rm Int70}(s)-12 \sqrt{2} \nu  \,{\rm Int71}(s)
\rrrlll
-6 \sqrt{2} \,{\rm Int80}(s)
+3\sqrt{2} \nu  \,{\rm Int81}(s)+6 \sqrt{2} \,{\rm Int9}(s)\right)
\right]\right\} n_{\la i}n_{j}n_{k \ra},\\
%%%%%%%%%%%%%
I_3&=\frac{G^3 m^4  \nu}{z^5} \left\{-\frac{1652}{3}+1652 \nu +(140-420 \nu ) \ln(8 \gamma)
+s^2 \left(\frac{1888}{5}
-\frac{5664 \nu }{5}+(-96+288 \nu ) \ln(8 \gamma)\right)
\rl+s^5 \left(\frac{2596}{15}-\frac{2596}{5}\nu-8\,{\rm Int11}(s)+24 \nu  \,{\rm Int11}(s)
+(-44+132 \nu ) \ln(8 \gamma)\right)\right\}n_{\la i}n_{j}n_{k} n_{l \ra},
\end{align}
\ese
where, Int1(s), Int20(s), Int21(s),Int30(s), Int31(s),Int4(s), Int5(s),Int6(s), Int70(s), Int71(s), 
Int80(s), Int81(s),Int9(s), Int10(s) read 
\bse\label{numints}
\begin{align}
{\rm Int1}(s)&=4\,\int_{s}^{1} dy\left({5-3y\over y^{5}}\right)\ln\left(s^{-3/2}\left(g(s)-g(y)\right)\right),\\
{\rm Int20}(s)&=\int_{s}^{1} dy\left({1540-1876 y+522 y^2\over 7 y^{6}}\right)\ln\left(s^{-3/2}\left(g(s)-g(y)\right)\right),\\
{\rm Int21}(s)&=\int_{s}^{1} dy\left({1365-2296 y+831 y^2\over 7 y^{6}}\right)\ln\left(s^{-3/2}\left(g(s)-g(y)\right)\right),\\
{\rm Int30}(s)&=4\,\int_{s}^{1} dy\left(\frac{(5-3 y)(5-y)}{y^6}\right)\ln\left(s^{-3/2}\left(g(s)-g(y)\right)\right),\\
{\rm Int31}(s)&=2\,\int_{s}^{1} dy\left(\frac{(5-3 y)(5-9 y)}{y^6}\right)\ln\left(s^{-3/2}\left(g(s)-g(y)\right)\right),\\
{\rm Int4}(s)&=4\,\int_{s}^{1} dy\left({5-3y\over y^{5}}\right)\left(\frac{h_0(s)-h_0(y)}{g(s)-g(y)}\right),\\
{\rm Int5}(s)&=4\,\int_{s}^{1} dy\left({5-3y\over y^{5}}\right)\left(\frac{h_1(s)-h_1(y)}{g(s)-g(y)}\right),\\
{\rm Int6}(s)&=\int_{s}^{1} dy\left({7-6y\over y^{9/2} \sqrt{1-y}}\right)\ln\left(s^{-3/2}\left(g(s)-g(y)\right)\right),\\
{\rm Int70}(s)&=\int_{s}^{1} dy\left({4675-3395 y-1548 y^2+684 y^3\over 18 y^{13/2} \sqrt{1-y}}\right)\ln\left(s^{-3/2}\left(g(s)-g(y)\right)\right),\\
{\rm Int71}(s)&=\int_{s}^{1} dy\left({1100+35 y-2133 y^2+1044 y^3\over 18 y^{13/2} \sqrt{1-y}}\right)\ln\left(s^{-3/2}\left(g(s)-g(y)\right)\right),\\
{\rm Int80}(s)&=\int_{s}^{1} dy\left(\frac{(5-y)(7-6 y)}{y^{11/2} \sqrt{1-y}}\right)\ln\left(s^{-3/2}\left(g(s)-g(y)\right)\right),\\
{\rm Int81}(s)&=\int_{s}^{1} dy\left(\frac{(5-9 y)(7-6 y)}{y^{11/2} \sqrt{1-y}}\right)\ln\left(s^{-3/2}\left(g(s)-g(y)\right)\right),\\
{\rm Int9}(s)&=\int_{s}^{1} dy\left({7-6y\over y^{9/2} \sqrt{1-y}}\right)\left(\frac{h_0(s)-h_0(y)}{g(s)-g(y)}\right),\\
{\rm Int10}(s)&=\int_{s}^{1} dy\left({7-6y\over y^{9/2} \sqrt{1-y}}\right)\left(\frac{h_1(s)-h_1(y)}{g(s)-g(y)}\right),\\
{\rm Int11}(s)&=\int_{s}^{1} dy\left({175-72 y^2\over y^{6}}\right)\ln\left(s^{-3/2}(g(s)-g(y))\right).
\end{align}
\ese
Using the above in Eq.~\eqref{lmf-hered-IJ}, performing contraction of indices and truncating the resulting 
expression at the 2.5PN order, we can now write the total hereditary contribution at 2.5PN order, 
solely expressed as a function of our PN parameter $\gamma$ and it reads
\begin{align}\label{heredlmf-FS}
\left({\mathcal{F}_{P}^{i}}\right)_{\rm hered}&=
-{32\over105}{c^4\over G}\gamma^{11/2}\sqrt{1-4\nu}\nu^2
\left[
\gamma ^{3/2} \left(s \left(\frac{221}{10}-9 \ln(8 \gamma)\right)
+s^2 \left(-\frac{304}{15}+8 \ln(8 \gamma)\right)\rrll
+4\sqrt{1-s} s^{9/2} {\rm Int6}(s)
+s^5 \left(-\frac{11}{6}+2 \,{\rm Int1}(s)+\ln(8 \gamma)\right)
\right)
\rl+\gamma^{5/2}\left(-\frac{89335}{216}+\frac{5255}{54}\nu+\frac{5525}{36} \ln(8 \gamma)-\frac{325}{9} \nu  \ln(8 \gamma)
+s \left(\frac{30893}{135}+\frac{41969}{270}\nu-\frac{682}{9} \ln(8 \gamma)-\frac{563}{9} \nu  \ln(8 \gamma)\right)
\rrll+s^2 \left(\frac{103237}{270}-\frac{268487}{540}\nu-\frac{1399}{9} \ln(8 \gamma)+\frac{3617}{18} \nu  \ln(8 \gamma)\right)
+s^3 \left(-\frac{9584}{45}+\frac{2224}{9}\nu+\frac{256}{3} \ln(8 \gamma)-\frac{304}{3} \nu  \ln(8 \gamma)\right)
\rrll+s^4 \left(\frac{4675}{216}-\frac{275}{54}\nu-\frac{425}{18}\,{\rm Int1}(s)+\frac{50}{9} \nu  \,{\rm Int1}(s)-\frac{425}{36} \ln(8 \gamma)
+\frac{25}{9} \nu\ln(8 \gamma)\right)
\rrll+s^5 \left(\frac{55}{54}-\frac{275}{54}\nu-\frac{10}{9}\,{\rm Int1}(s)+\frac{50}{9} \nu  \,{\rm Int1}(s)-\frac{5}{9} \ln(8 \gamma)
+\frac{25}{9} \nu \ln(8 \gamma)\right)
\rrll+s^6 \left(-\frac{1969}{270}+\frac{451}{60}\nu+\frac{32}{3}\,{\rm Int1}(s)-\frac{37}{3} \nu  \,{\rm Int1}(s)-\frac{8}{63}\,{\rm Int11}(s)
+\frac{8}{21} \nu \,{\rm Int11}(s)-2 \,{\rm Int20}(s)+2 \nu  \,{\rm Int21}(s)\rrrlll+\,{\rm Int30}(s)-\nu  \,{\rm Int31}(s)-\,{\rm Int4}(s)
+\frac{1}{2} \nu  \,{\rm Int5}(s)+\frac{43}{9}\ln(8 \gamma)-\frac{13}{2} \nu  \ln(8 \gamma)\right)
\rrll+\frac{1}{\sqrt{1-s}}\left(s^{9/2} \left(-\frac{122}{9}  \,{\rm Int6}(s)+\frac{59}{3}  \nu  \,{\rm Int6}(s)\right)
\rrrlll+s^{11/2} \left( \nu \,{\rm Int10}(s)+\frac{296}{9}  \,{\rm Int6}(s)-\frac{122}{3}  \nu  \,{\rm Int6}(s)
-4  \,{\rm Int70}(s)+4  \nu  \,{\rm Int71}(s)+2 \,{\rm Int80}(s)\rrrrllll- \nu  \,{\rm Int81}(s)
-2  \,{\rm Int9}(s)\right)+s^{13/2} \left(- \nu  \,{\rm Int10}(s)-\frac{58}{3} \,{\rm Int6}(s)
+21  \nu  \,{\rm Int6}(s)+4  \,{\rm Int70}(s)\rrrrllll-4  \nu  \,{\rm Int71}(s)-2  \,{\rm Int80}(s)
+ \nu \,{\rm Int81}(s)+2  \,{\rm Int9}(s)\right)\right)
\right)
+{\cal O}\left({1\over c^6}\right)
\right]{n_i}.
\end{align}
Note again, that leading order hereditary contribution (1.5PN tail) is proportional to various 
powers of $s$ and hence would be absent when we specialize our result to case (b). The reason 
is similar to the one given at the end of Sec.~\ref{subsubsec:lmf-inst-FS} to explain the 
absence of the Newtonian terms in instantaneous part for case (b). Observe that, the first two terms of 
Eq.~\eqref{lmf-hered-IJ} are proportional to the $I_{ijk}^{(4)}$ and $I_{ijk}^{(6)}$, and these 
are the ones which should contributing at the 1.5PN order. But, since Newtonian order expression 
for $I_{ijk}^{(n)}$ vanishes for $n>2$, for the case of infall from infinity, there would be no 
contribution at the 1.5PN order for case (b).
\subsubsection{Case (b): Infall from infinity}
\label{subsubsec:lmf-hered-IS}
Using the argument, that the Newtonian order expression for $I_{ijk}^{(n)}$ vanishes 
for $n>2$ in the case of infall from infinity, in Eq.~\eqref{lmf-hered-IJ}, we can immediately see 
that only first two terms of Eq.~\eqref{lmf-hered-IJ} are going to contribute to the linear 
momentum flux. And thus we need to evaluate only the integrals appearing in these two terms. In this 
case, the relevant integrals take the following form    
\bse\label{heredint-IS}
\begin{align}
\label{heredint1-IS}
I_{1}&=\int_{-\infty}^{u} d\tau \left[\ln \left({c\over 2 z}(u-\tau)\right)+{11\over12} \right]
I^{(5)}_{ij} (\tau),\\
\label{heredint2-IS}
I_{2}&=\int_{-\infty}^{u} d\tau \left[\ln \left({c\over 2 z}(u-\tau)\right)+{97\over60} \right]
I^{(6)}_{ijk} (\tau).
\end{align}
\ese
With, required derivatives of the source multipole moments, expressed in terms of $z$, and the 
1PN trajectory (given by Eq.\eqref{1PNtraj-IS}) we can evaluate these integrals and they read
\bse\label{heredinteva-IS}
\begin{align}
I_{1}&=\frac{G^2 m^3  \nu}{z^4}\left(-\frac{71}{6}-\frac{5 \pi
   }{\sqrt{3}}-5 \ln\left[\frac{2 \gamma }{3}\right]+\gamma 
   \left(-\frac{2497}{21}+\frac{166 \pi }{\sqrt{3}}-\frac{2161
   }{42}\nu-22 \sqrt{3} \pi  \nu +34 (1- \nu ) \ln\left[\frac{2
   \gamma}{3}\right]\right)\right)n_{\la i} n_{j \ra},\\
I_2&=\frac{G^{5/2} m^{7/2} \nu }{z^{9/2}}\sqrt{1-4\nu}\left(-\frac{8755}{9 \sqrt{2}}-\frac{850}{3} \sqrt{\frac{2}{3}} \pi
   +\frac{1030 \sqrt{2} \nu }{9}+\frac{200}{3} \sqrt{\frac{2}{3}} \pi 
   \nu +\left(-\frac{850 \sqrt{2}}{3}+\frac{200 \sqrt{2}}{3}\nu\right)
   \ln\left[\frac{2 \gamma }{3}\right]\right)n_{\la i}n_{j} n_{k \ra}.
\end{align}
\ese
Using the above result in Eq.~\eqref{lmf-hered-IJ}, we can write the complete hereditary contribution at 2.5PN order, 
as a function of our PN parameter $\gamma$, and it reads
\begin{align}\label{heredlmf-IS}
\left({\mathcal{F}_{P}^{i}}\right)_{\rm hered}&=
-{32\over105}{c^4\over G}\gamma^{8}\sqrt{1-4\nu}\nu^2
\left[
\frac{65195}{216}+\frac{5525 \pi }{36
   \sqrt{3}}+\left(-\frac{3835}{54}-\frac{325 \pi }{9 \sqrt{3}}\right)
   \nu +\left(\frac{5525}{36}-\frac{325}{9}\nu\right)
   \ln\left[\frac{2 \gamma }{3}\right]
+{\cal O}\left({1\over c^6}\right)
\right]{n_i}.
\end{align}
\subsection{Total Linear Momentum Flux}
\label{subsec:lmfhoc-total}
\subsubsection{Case (a): Infall from a finite distance}
\label{subsubsec:lmfhoc-total-FS}
For this case, Eq.~\eqref{instlmf-FS} and Eq.~\eqref{heredlmf-FS} can be added to write the complete 2.5PN accurate 
expression for the linear momentum flux, expressed as a function of the parameter $\gamma$, and it reads
\begin{align}\label{totallmf-FS}
{\mathcal{F}_{P}^{i}}&=-{32\,\sqrt{2}\over105}{c^4\over G}\sqrt{1-s}\gamma^{11/2}\sqrt{1-4\nu}\nu^2
\left[s
%%%%%%
+\gamma\left(
-\frac{425}{36}+\frac{25}{9}\nu
+s\left(-\frac{71}{18}+\frac{277}{36}\nu\right)
+s^2 \left(\frac{61}{6}-\frac{113}{12}\nu\right)
\right)\rl
%%%%%%
+\frac{\gamma ^{3/2}}{\sqrt{2} \sqrt{1-s}}
\left(
s \left(\frac{221}{10}-9 \ln(8 \gamma)\right)
+s^2 \left(-\frac{304}{15}+8 \ln(8 \gamma)\right)
+4 \sqrt{1-s} s^{9/2} \,{\rm Int6}(s)
+s^5 \left(-\frac{11}{6}+2\,{\rm Int1}(s)+\ln(8 \gamma)\right)
\right)
\rl
%%%%%%
+\gamma ^2 \left(
\frac{363379}{2376}-\frac{315163}{1584}\nu+\frac{14635}{396}\nu^2
+s\left(-\frac{99647}{594}+\frac{278611}{1584}\nu+\frac{12965}{3168}\nu^2\right)
+s^2 \left(-\frac{4801}{132}+\frac{125819}{792}\nu-\frac{129959}{1584}\nu^2\right)
\rrll+s^3 \left(\frac{7399}{264}-\frac{12527}{132}\nu+\frac{13873}{352}\nu^2\right)
\right)
%%%%%%
+{\gamma^{5/2}\over \sqrt{2}\sqrt{1-s}}\left(
-\frac{89335}{216}+\frac{31339}{270}\nu+\frac{5525}{36} \ln(8 \gamma)-\frac{325}{9} \nu  \ln(8 \gamma)
\rrll+s \left(\frac{30893}{135}+\frac{32321}{270}\nu-\frac{682}{9} \ln(8 \gamma)-\frac{563}{9} \nu  \ln(8 \gamma)\right)
+s^2 \left(\frac{103237}{270}-\frac{253463}{540}\nu-\frac{1399}{9} \ln(8 \gamma)+\frac{3617}{18} \nu  \ln(8 \gamma)\right)
\rrll+s^3 \left(-\frac{9584}{45}+\frac{1184}{5} \nu+\frac{256}{3} \ln(8 \gamma)-\frac{304}{3} \nu  \ln(8 \gamma)\right)
+s^4 \left(\frac{4675}{216}-\frac{275}{54}\nu-\frac{425}{18}\,{\rm Int1}(s)+\frac{50}{9} \nu  \,{\rm Int1}(s)-\frac{425}{36} \ln(8 \gamma)
\rrrlll+\frac{25}{9} \nu\ln(8 \gamma)\right)
+s^5 \left(\frac{55}{54}-\frac{275}{54}\nu-\frac{10}{9}\,{\rm Int1}(s)+\frac{50}{9} \nu  \,{\rm Int1}(s)-\frac{5}{9} \ln(8 \gamma)
+\frac{25}{9} \nu \ln(8 \gamma)\right)
\rrll+s^6 \left(-\frac{1969}{270}+\frac{451}{60}\nu+\frac{32}{3}\,{\rm Int1}(s)-\frac{37}{3} \nu  \,{\rm Int1}(s)-\frac{8}{63}\,{\rm Int11}(s)
+\frac{8}{21} \nu \,{\rm Int11}(s)-2 \,{\rm Int20}(s)+2 \nu  \,{\rm Int21}(s)\rrrlll+\,{\rm Int30}(s)-\nu  \,{\rm Int31}(s)-\,{\rm Int4}(s)
+\frac{1}{2} \nu  \,{\rm Int5}(s)+\frac{43}{9}\ln(8 \gamma)-\frac{13}{2} \nu  \ln(8 \gamma)\right)
\rrll+\sqrt{1-s}\left(s^{9/2} \left(-\frac{122}{9}\,{\rm Int6}(s)+\frac{59}{3} \nu  \,{\rm Int6}(s)\right)
+s^{11/2} \left(\nu \,{\rm Int10}(s)+\frac{58}{3}\,{\rm Int6}(s)-21 \nu\,{\rm Int6}(s)-4\,{\rm Int70}(s)+4 \nu \,{\rm Int71}(s)
\rrrrllll+2 \,{\rm Int80}(s)-\nu  \,{\rm Int81}(s)-2\,{\rm Int9}(s) \right)
\right)
\right)
+{\cal O}\left({1\over c^6}\right)
\right]{n_i}.\end{align}
\subsubsection{Case (b): Infall from infinity}
\label{subsubsec:lmfhoc-total-IS}
For this case, Eq.~\eqref{instlmf-IS} and Eq.~\eqref{heredlmf-IS} can be added to get the complete 2.5PN accurate 
expression for the linear momentum flux, in harmonic coordinates, expressed as a function of the 
parameter $\gamma$, and it reads
\begin{align}\label{totallmf-IS}
{\mathcal{F}_{P}^{i}}&=-{32\,\sqrt{2}\over105}{c^4\over G}\sqrt{1-s}\gamma^{11/2}\sqrt{1-4\nu}\nu^2
\left[
%%%%%%
\gamma\left(
-\frac{425}{36}+\frac{25}{9}\nu
\right)
%%%%%%
+\gamma ^2 \left(
\frac{363379}{2376}-\frac{315163}{1584}\nu+\frac{14635}{396}\nu^2
\right)\rl
%%%%%%
+\gamma^{5/2}\left(
\frac{65195}{216 \sqrt{2}}+\frac{5525 \pi }{36
   \sqrt{6}}+\left(-\frac{14111}{270 \sqrt{2}}-\frac{325 \pi }{9
   \sqrt{6}}\right) \nu +\left(\frac{5525}{36 \sqrt{2}}-\frac{325}{9
   \sqrt{2}}\nu\right) \ln\left[\frac{2 \gamma }{3}\right]
\right)
+{\cal O}\left({1\over c^6}\right)
\right]{n_i}.
\end{align}
%%%%%%%%%%%%%%%%%%%%%%%%%%%%%%%%%%%%%%%%%%%%%%Recoil-Velocity%%%%%%%%%%%%%%%%%%%%%%%%%%%%%%%%%%%%%%
\section{Recoil Velocity}
\label{sec:recvel}
With, the 2.5PN expression for linear momentum flux emitted during the radial infall of two compact 
objects for two different situations (case (a) and case (b)), in harmonic coordinates, we can now 
use the momentum balance argument to write the loss rate of linear momentum from the source (through 
outgoing gravitational waves) and it reads
\be
\label{mombal}
{dP^i\over du}=-{\mathcal F}_P^i(u).
\ee
The net loss of linear momentum can be obtained by integrating the balance equation, i.e.
\be
\label{deltamom}
{\Delta P^i}=-\int_{-\infty}^{u}\,du'\,{\mathcal F_P^i(u')}.
\ee
\subsection{Case (a): Infall from a finite distance}
\label{subsec:recvel-FS}
In this case, Eq.~\eqref{deltamom} can be written as
\begin{align}
\label{deltamom-FS}
{\Delta P^i}&=-\int_{u(z_i)}^{u(z_f)}\,du\,{\mathcal F_P^i(u)} \nonumber\\ &
             =-\int_{z_i}^{z_f}\,{dz\over \dot{z}(z)}\,{\mathcal F_P^i(z)} \nonumber\\ &
             ={G m\over c^2}\int_{\gamma_i}^{\gamma_f}\,{d\gamma\over \gamma^2 \dot{z}(\gamma)}\,{\mathcal F_P^i(\gamma)}.
\end{align}
as $\gamma=(G m/c^2 z)$ and $dz=-(G\,m/c^2\,\gamma^2)d\gamma$. Here, $z_f$ denotes some final
separation where we would like terminate our integral. Also, two limiting values of the parameter, 
$\gamma$, are $\gamma_i=(G m/c^2 z_i)$ and $\gamma_f=(G m/c^2 z_f)$.

We can use the 2.5PN expressions for the linear momentum flux (Eq.~\eqref{totallmf-FS}) and for 
$\dot{z}$ (Eq.~\eqref{zdot-FS}) in the above integral to compute the total loss of linear 
momentum from the source during the radial infall from an initial separation of $z_i$ ($\gamma_i$) 
to a final separation of $z_f$ ($\gamma_f$).  Since, linear momentum flux given by 
Eq.~\eqref{totallmf-FS} involves some integrals (Eq.~\eqref{numints}) which have to be computed 
numerically, we can not have an analytical expression for the total loss of the linear momentum 
from the source and thus need to be computed numerically. The corresponding recoil velocity 
can be computed as  
\be\label{recvel-FS}
\Delta V^i= \Delta P^i/m
\ee
where, m is the total mass of the system.
We shall present our estimates for the recoil velocity for the case of infall from a finite distance  
in the next section where we shall discuss all our findings.
\subsection{Case (b): Infall from infinity}
In this case, the loss of linear momentum can be given by the integral
\begin{align}
\label{deltamom-IS}
{\Delta P^i}&=-\int_{-\infty}^{u(z_f)}\,du\,{\mathcal F_P^i(u)} \nonumber\\ &
             =-\int_{\infty}^{z_f}\,{dz\over \dot{z}(z)}\,{\mathcal F_P^i(z)} \nonumber\\ &
             ={G m\over c^2}\int_{0}^{\gamma_f}\,{d\gamma\over \gamma^2 \dot{z}(\gamma)}\,{\mathcal F_P^i(\gamma)}.
\end{align}
The 2.5PN expressions for the linear momentum flux (Eq.~\eqref{totallmf-IS}) and for $\dot{z}$
(Eq.~\eqref{zdot-IS}) can be used in the above to compute the total loss in the linear momentum during the 
radial infall of the two objects for the case of infall from infinity. Next, Eq.~\eqref{recvel-FS} can be used 
to compute the corresponding expression for the recoil velocity. We find for the 2.5 PN recoil 
velocity, in harmonic coordinates, expressed in terms of $\gamma$ as
\begin{align}
\label{recvel-IS}
\Delta V^i&={16\over 105}{c\gamma_f^4\sqrt{1-4\nu}\nu^2}
\left[\gamma_f\left(-\frac{85}{18}+\frac{10}{9}\nu \right)
+\gamma_f^2\left(\frac{146627}{3564}-\frac{23399}{396}\nu+\frac{1105}{99}\nu^2\right)
\rl+\gamma_f^{5/2}\left(\frac{60095}{702 \sqrt{2}}+\frac{425 \pi }{9 \sqrt{6}}+\frac{425
}{9 \sqrt{2}}\ln\left[\frac{2 \gamma_f}{3}\right]+\nu 
\left(-\frac{12611 \sqrt{2}}{1755}-\frac{50}{9} \sqrt{\frac{2}{3}} \pi
-\frac{50}{9} \sqrt{2} \ln\left[\frac{2 \gamma_f}{3}\right]\right)\right)
+{\cal O}\left({1\over c^6}\right)
\right]{n_i}.
\end{align}       
%%%%%%%%%%%%%%%%%%%%%%%%%%%%%%%%%%%%%%%%%%%%%%DISCUSSION%%%%%%%%%%%%%%%%%%%%%%%%%%%%%%%%%%%%%%%%%%%
\section{Discussions and Conclusions}
\label{sec:results}
\begin{figure}
\includegraphics[width=0.502\textwidth,angle=0]{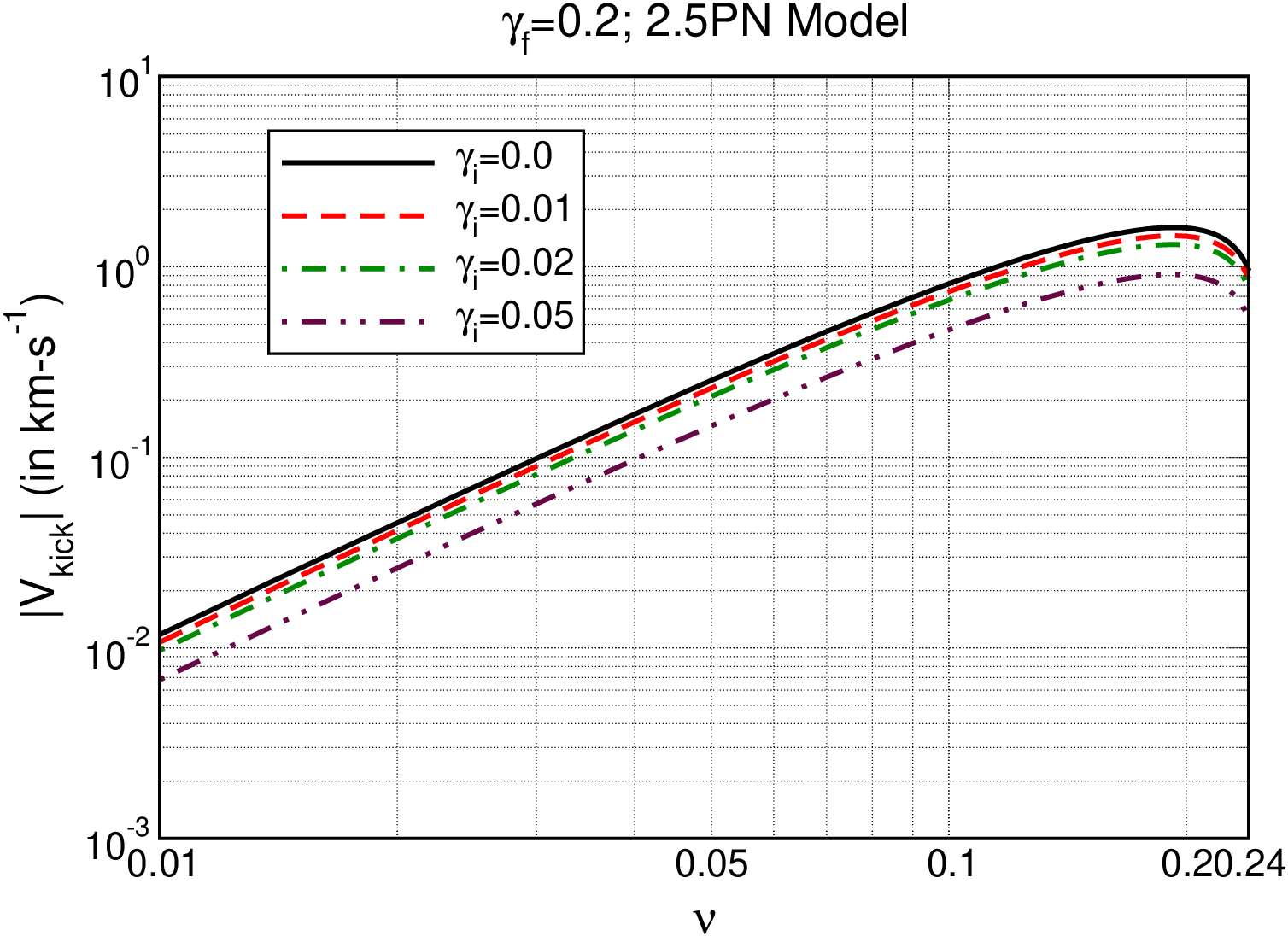}
\includegraphics[width=0.492\textwidth,angle=0]{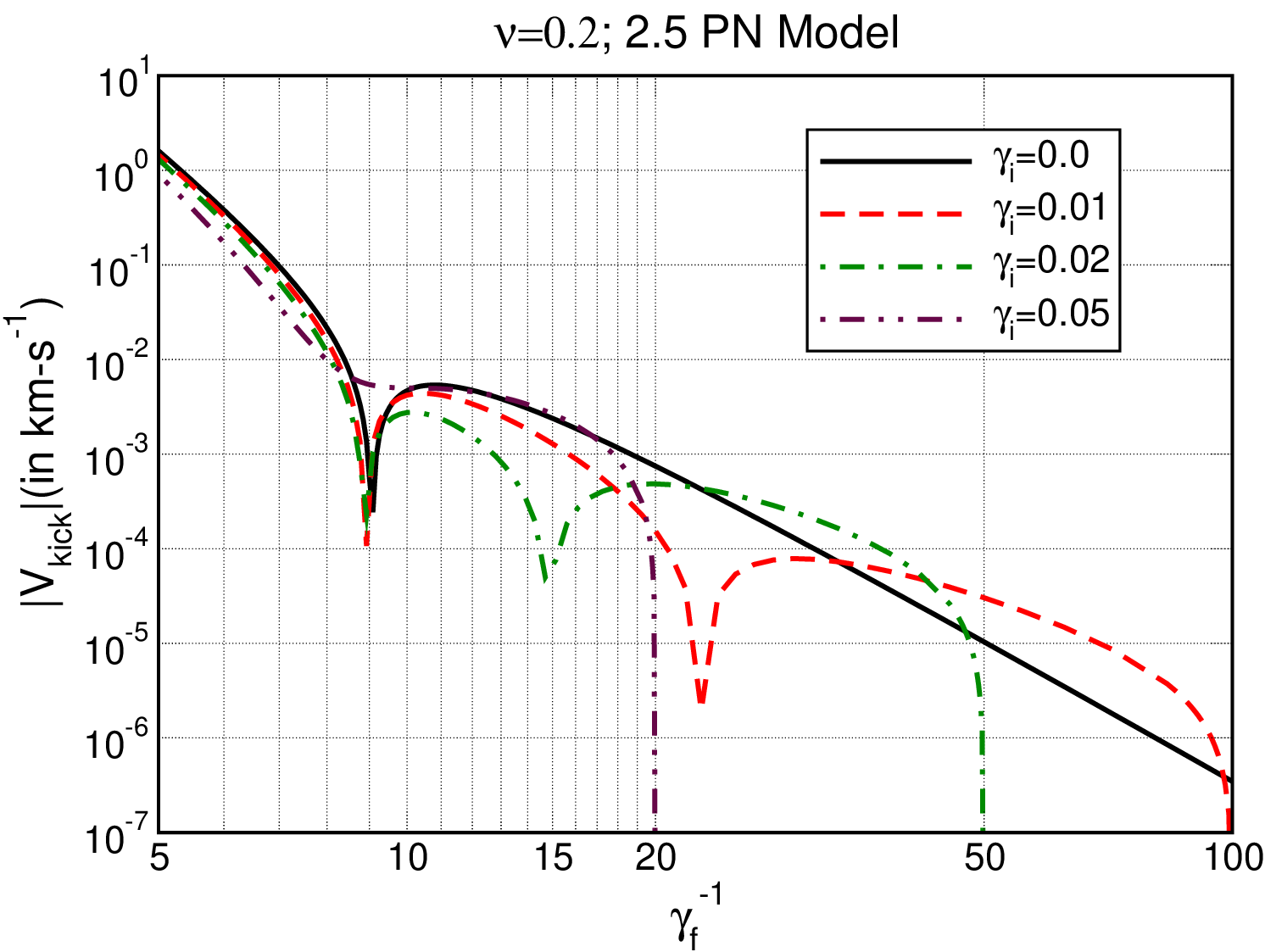}
\caption{Recoil velocity as a function of the mass parameter $\nu$ (left panel) and as a 
function of the post-Newtonian parameter $\gamma_f$ (right panel) has been plotted. The parameter, 
$\nu$, is known as symmetric mass ratio of the binary; the parameters $\gamma_f=(G\,m/c^2\,z_f)$ 
and $\gamma_i=(G\,m/c^2\,z_i)$ are the post-Newtonian parameters characterizing the final and 
initial separation of the two objects, respectively. For the plot in the left panel, 
the value of the parameter $\gamma_f$ has been fixed to 0.2, which corresponds to the final 
separation of 5 $G\,m/c^2$ between the two objects and then the recoil velocity 
as a function of the parameter $\nu$ has been plotted. Similarly, for the right panel, the value of the 
parameter $\nu$ has been fixed to 0.2 and recoil velocity as a function of the parameter $\gamma_f$ has 
been shown. These plots (both in the left and the right panel) also compare recoil velocity estimates 
for four different situations based on the binary's initial separation: $\gamma_i$=0.01, 0.02, 0.05, 
and 0.0 which correspond to the initial separation of the two objects of 100 $G\,m/c^2$, 50 $G\,m/c^2$, 
20 $G\,m/c^2$, and $\infty$ (infinite initial separation case), respectively.} 
\label{fig:recvel-vsnu-vsgf}
\end{figure}
\begin{figure}
\includegraphics[width=0.45\textwidth,angle=0]{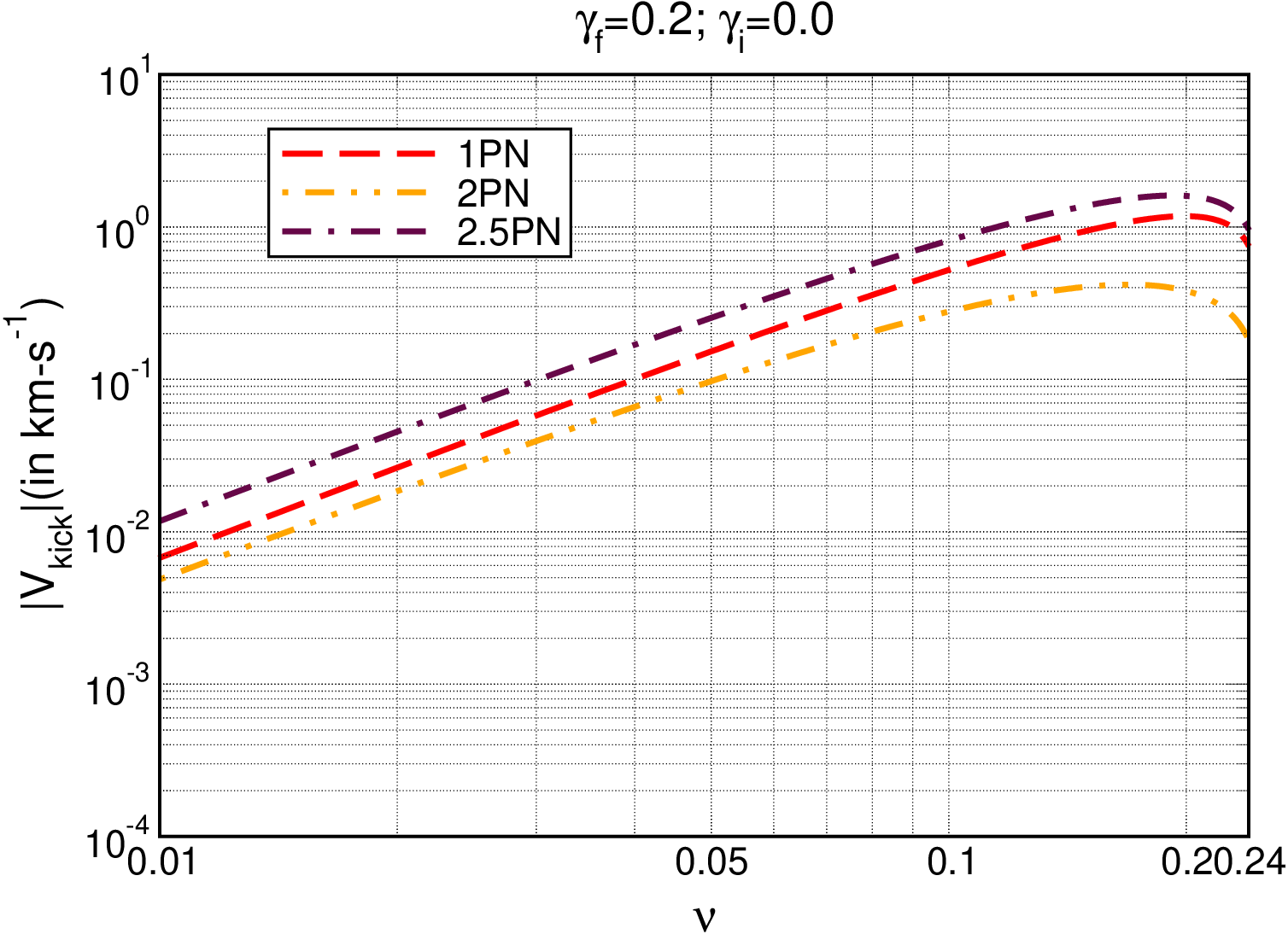}
%\hskip 0.2cm
\includegraphics[width=0.45\textwidth,angle=0]{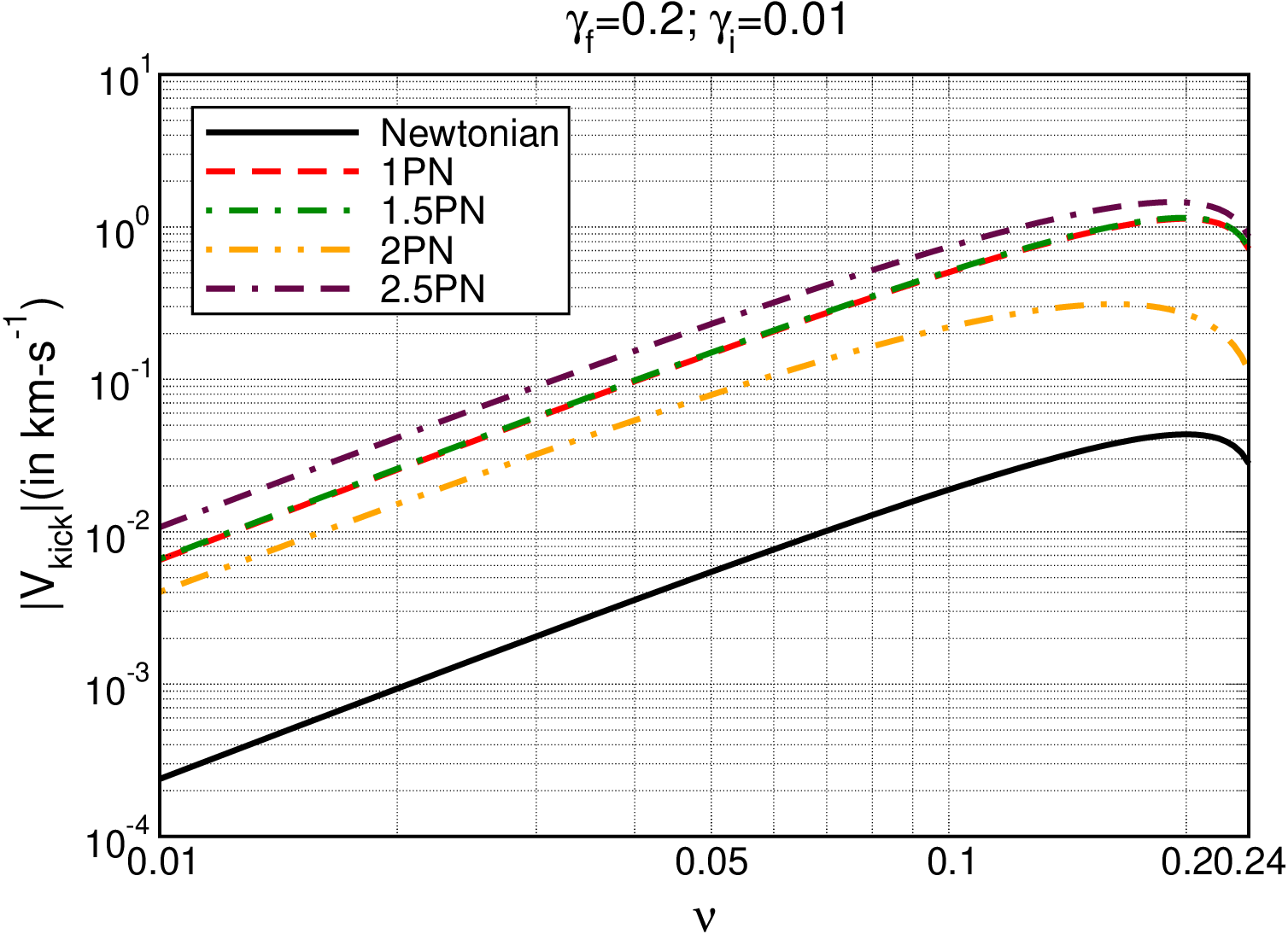}
%\vskip 0.2cm
\includegraphics[width=0.45\textwidth,angle=0]{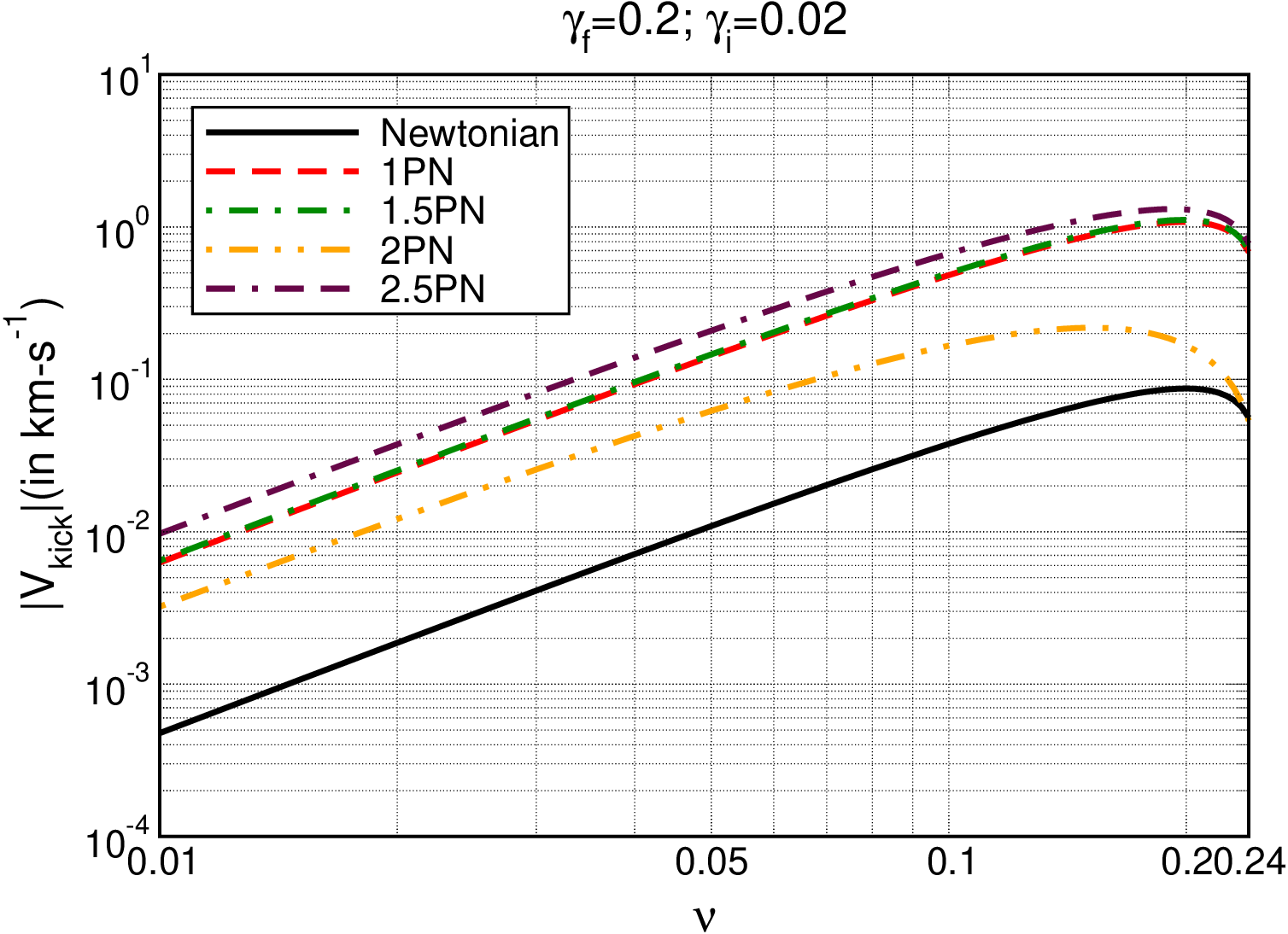}
%\hskip 0.2cm
\includegraphics[width=0.45\textwidth,angle=0]{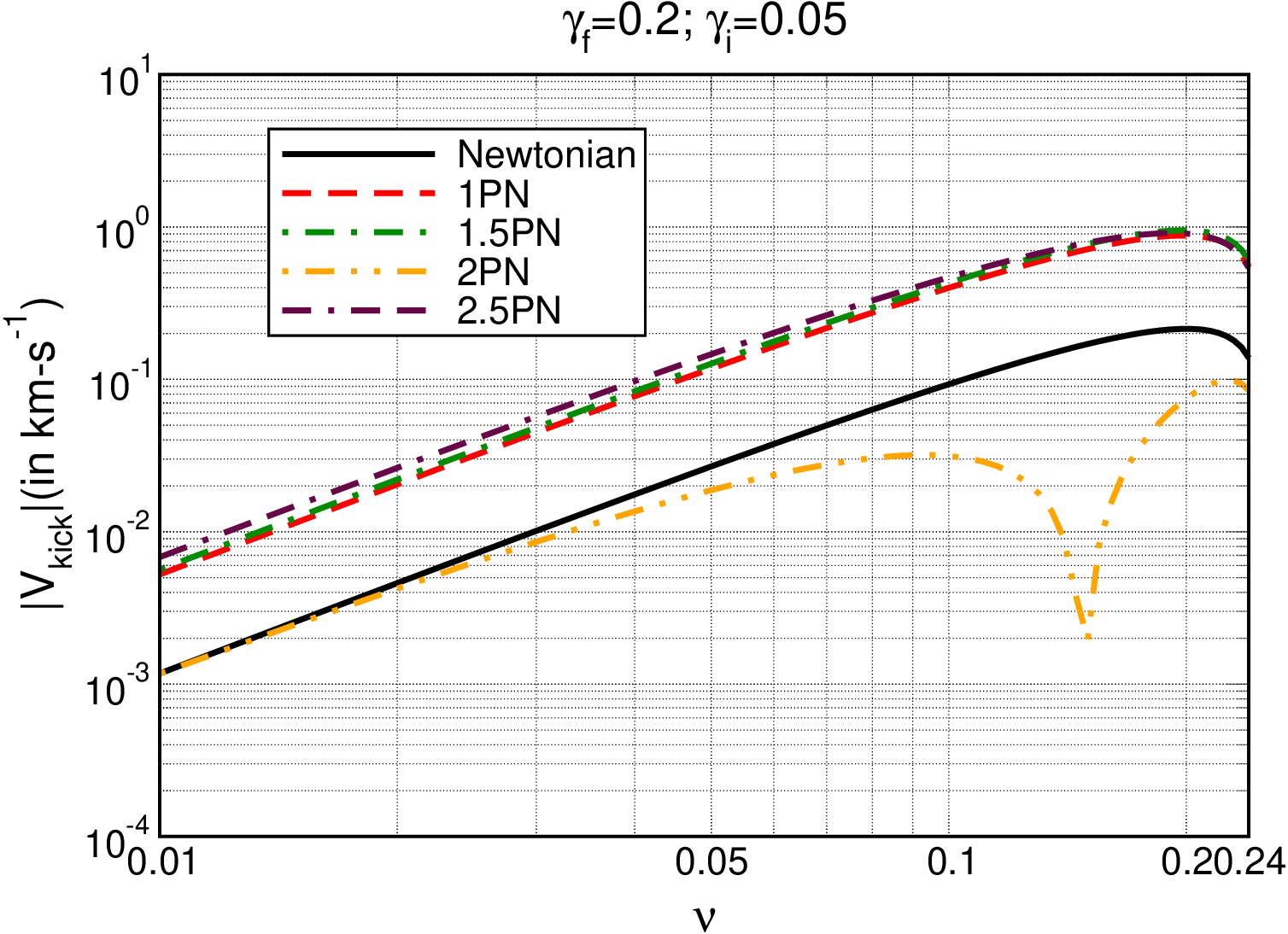}
\caption{Recoil velocity as a function of the parameter $\nu$ has been 
shown. For all the plots, the value of the parameter $\gamma_f$ has been 
fixed to 0.2 (which corresponds to the final separation of 5 $G\,m/c^2$ 
between the two objects under the radial infall). Plots in different panels 
also compare the results with different PN accuracy for four different 
situations: $\gamma_i$=0.01, 0.02, 0.05, and 0.0 which correspond to the initial 
separation (of the two objects in the problem) of 100 $G\,m/c^2$, 50 $G\,m/c^2$, 
20 $G\,m/c^2$, and $\infty$ (infinite initial separation case), respectively.}
\label{fig:recvelvsnu-nPN}
\end{figure}
The 2.5PN accurate expressions for the linear momentum flux emitted during the radial infall of two 
compact objects for two different situations (infall from some finite initial separation and infall 
from infinity), in harmonic coordinates, expressed in terms of the post-Newtonian parameter $\gamma$
(related to the separation of the two objects), has been given by Eq.~\eqref{totallmf-FS} and 
Eq.\eqref{totallmf-IS}. Next, we use these expressions to compute the associated recoil velocity of 
the source. Equation~\eqref{recvel-IS} gives the 2.5PN accurate analytical formula for the 
recoil velocity accumulated till any epoch during the binary's evolution (within the validity of PN 
approximations), for the case of infall from infinity, and can be used to compute related numerical 
estimates for the recoil velocity. Since linear momentum flux formula (Eq.~\eqref{totallmf-FS}), 
for the case which assumes the infall from some finite initial separation, involves some integrals 
(Eq.~\eqref{numints}) which can only be evaluated numerically, it is not possible to give analytical 
PN expressions for the accumulated recoil velocity for this case. Figures~\ref{fig:recvel-vsnu-vsgf} 
and \ref{fig:recvelvsnu-nPN} show the numerical estimates for the recoil velocity accumulated during 
the radial infall of two compact objects and we shall discuss them one by one. 

Figure~\ref{fig:recvel-vsnu-vsgf} plots recoil velocity as a function of $\nu$ (left panel) and as a 
function of the parameter $\gamma_f$ (right panel). Here, $\gamma_f$ is our post-Newtonian parameter
given by $\gamma_f=(G\,m/c^2\,z_f)$. For the plots in the left panel of Fig.~\ref{fig:recvel-vsnu-vsgf}
the value of the parameter $\gamma_f$ has been fixed to 0.2 and then the recoil velocity has been 
plotted as a function of $\nu$ for the range of $\nu=0.01$ (nearly test particle limit) to $\nu=0.24$ 
(nearly symmetric binary). The right panel shows the variations in recoil velocity estimates as 
a function of the parameter $\gamma_f$ for a range of values between $\gamma_f=0.01$ to $\gamma_f=0.2$, 
for a binary with $\nu=0.2$. These plots (both in the right and the left panel) also compare 
the recoil velocity estimates for four different situations related to the initial separation of the 
two objects under the radial infall. The recoil velocity estimates have been plotted for four different 
values of the parameter $\gamma_i=(G\,m/c^2\,z_i)$: $\gamma_i$=0.01, 0.02, 0.05 and 0.0 which correspond 
to the initial separation of 100 $G\,m/c^2$, 50 $G\,m/c^2$, 20 $G\,m/c^2$, and $\infty$ (infinite 
initial separation case), respectively.   

Based on the estimates shown in the left panel of Fig.~\ref{fig:recvel-vsnu-vsgf}, we find that the 
recoil velocity is maximum for a binary with $\nu \sim 0.19$ and is of the order of 
$\sim$ 1.6${\rm km\,s}^{-1}$. Also, the behavior of the plots is as one would expect: recoil velocity   
is maximum for the infinite initial separation case and estimates become smaller for situations which 
assume that infall shall proceed from smaller separations.\footnote{Note that for finite separation 
cases ($\gamma_i$=0.01, 0.02, 0.05), initially the contribution exceeds as compared to the case of 
infinite initial separation ($\gamma_i$=0.0): this is not surprising since this contribution comes 
from the Newtonian terms which are absent in infinite initial separation case.} However, we observe that estimates for the 
recoil velocity for all four situations ($\gamma_i$=0.01, 0.02, 0.05 and 0.0) are of the same order, 
indicating that most of the contribution comes from late stages of the infall. 

Although, we are not 
aware of a study which provides recoil velocity accumulated only during the premerger phase of a binary 
under the radial infall, a comparison with some other analytical/numerical work (which also involve 
contributions from the merger phase of the binary evolution) will be useful. For our purpose 
(head-on collision of two 
nonspinning compact objects), closest comparisons can be made using the results of \cite{Choi:2007eu} 
(Numerical Relativity) 
and of \cite{Nakamura:1983hk} (black hole perturbation theory). As compared to the recoil velocity 
estimates of about 2-5 ${\rm km\,s}^{-1}$ of \cite{Choi:2007eu} for a black hole binary (with $\nu=0.24$) 
under radial infall, our estimates using (Eq.~\eqref{recvel-IS}) suggest a recoil velocity of the 
order of 0.95 ${\rm km\,s}^{-1}$ for the same system ({\it i.e.} with $\nu=0.24$). Reference
\cite{Nakamura:1983hk} suggests that the recoil velocity accumulated during the head-on 
infall and plunge of a test particle in to a Schwarzschild black hole is given by 
$\Delta V=8.73\times10^{-4}\,\nu\,c$, which, compared to our estimates of the recoil velocity using 
the test particle limit of Eq.~\eqref{recvel-IS} ($\Delta V=4.06\times10^{-4}\,\nu\,c$), is larger 
by a factor of two. The difference between our estimates and other related estimates is 
possibly due to the fact that we do not evolve our system till it merges.    

Figure~\ref{fig:recvelvsnu-nPN} plots the recoil velocity as a function of $\nu$. 
For all the plots, the value of the parameter $\gamma_f$ has been fixed to 0.2. Four 
panels correspond to the four initial separations which have been discussed above while describing 
Figure~\ref{fig:recvelvsnu-nPN}. Each panel compares the recoil velocity estimates using results 
with different PN accuracy (Newtonian, $\cdots$, 2.5PN). It should be noted that we are terminating 
all our computations at 
$\gamma_f=0.2$ ({\it i.e} when the distance between the two objects is 5 $G\,m/c^2$). The reason for 
this is related to the validity of our formulas beyond this final separation. Generally, it is 
believed that when higher order PN corrections start becoming comparable to the leading order 
contribution in the series and such a series becomes less reliable. A few checks with our analytical  
expressions indicate that these estimates are reliable for separations larger than 5 $G\,m/c^2$ 
($\gamma=0.2$) and this is why we terminate all our computations at this value ($\gamma=0.2$). 
%%%%%%%%%%%%%%%%%%%%%%%%%%%%%%%%%%%%%%%%%%%%%%%%%%%%%%%%%%%%%%%%%%%%%%%%%%%%%%%%%%%%%%%%%%%%%%%%%%%%%%%%%%
%%%%%%%%%%%%%%%%%%%%%%%%%%%%%%%%%%%%%%%%%%%%%%%%%%%%%%%%%%%%%%%%%%%%%%%%%%%%%%%%%%%%%%%%%%%%%%%%%%%%%%%%%%
\appendix
\section{The 2.5PN linear momentum flux and recoil velocity in ADM coordinates}
\label{app:admlmf}
In the above, we have given the 2.5PN accurate analytical expression for the linear momentum flux due to 
radial infall of two compact objects under mutual gravitational influence, in harmonic coordinates. In this 
section we shall provide equivalent formulas in ADM coordinates.    
\subsection{Case (a): Infall from a finite distance}
The 2.5PN accurate analytical expression for the linear momentum flux in ADM coordinates can be obtained by 
using the following relation
\be\label{lmfADM}
({\mathcal{F}_{P}^{i}})_{\rm ADM}={\mathcal{F}_{P}^{i}}+\delta_{({\rm Har} \rightarrow {\rm ADM})}{\mathcal{F}_{P}^{i}}.
\ee
Here, ${\mathcal{F}_{P}^{i}}$ is given by Eq.~\eqref{totallmf-FS} and 
$\delta_{({\rm Har} \rightarrow {\rm ADM})}{\mathcal{F}_{P}^{i}}$ reads
\be\label{deltaHar2ADM-FS}
\delta_{({\rm Har} \rightarrow {\rm ADM})}{\mathcal{F}_{P}^{i}}=
-\frac{32\,\sqrt{2}}{105}\frac{c^4}{G}\sqrt{1-s}\gamma ^{15/2}\sqrt{1-4 \nu}\nu ^2 
\left(-\frac{1}{4}+\frac{\nu }{2}
+s\left(\frac{5}{8}+\frac{9}{4}\nu\right)
+s^2\left(-\frac{1}{8}-\frac{3}{2}\nu\right)
+s^3 \left(-\frac{3}{8}-\frac{9}{2}\nu\right)
\right)n_i.
\ee 
\subsection{Case (b): Infall from infinity}    
In this case, expression for the linear momentum flux in ADM coordinates can be obtained using 
Eq.~\eqref{lmfADM}, with ${\mathcal{F}_{P}^{i}}$ given by Eq.~\eqref{totallmf-IS} and 
$\delta_{({\rm Har} \rightarrow {\rm ADM})}{\mathcal{F}_{P}^{i}}$ as
\be\label{deltaHar2ADM-IS}
\delta_{({\rm Har} \rightarrow {\rm ADM})}{\mathcal{F}_{P}^{i}}=
-\frac{32 \sqrt{2}}{105}{c^4 \over G}\gamma ^{15/2}\sqrt{1-4 \nu}\nu ^2  
\left(-\frac{1}{4}+\frac{\nu }{2}\right)n_i.
\ee 
In this case we can also write the recoil velocity expression in ADM coordinates by using the 
following relation
\be\label{recvelADM}
(\Delta{V}^i)_{\rm ADM}=\Delta{V}^i+\delta_{({\rm Har} \rightarrow {\rm ADM})}\Delta V^i.
\ee
Here, $\Delta{V}^i$ is given by Eq.~\eqref{recvel-IS} and $\delta_{({\rm Har} \rightarrow {\rm ADM})}\Delta V^i$ is given by
\be\label{deltarecvelHar2ADM-IS}
\delta_{({\rm Har} \rightarrow {\rm ADM})}V^i={16\over 105}{c\gamma_f^6\sqrt{1-4\nu}\nu^2}\left(-{1\over12}+{\nu\over 6}\right)n_i.
\ee
\begin{acknowledgments}
I thank Bala R. Iyer for suggesting this problem. I thank Bala R. Iyer and K. G. Arun for 
discussions and useful suggestions on the manuscript.
\end{acknowledgments}
\bibliography{ref-list}
\end{document}